\documentclass{aa}  

\usepackage{graphicx}
\usepackage{txfonts}
\usepackage{graphicx}
\usepackage{graphics}
\usepackage{subfig}
\usepackage{natbib}
\usepackage{rotating}
\usepackage{amsmath}
\usepackage{wasysym}
\begin{document} 

\newcommand{\sigmahtwo}{$\Sigma_{\rm H_2}$}
\newcommand{\sigmahi}{$\Sigma_{\rm HI}$}
\newcommand{\sigmasfr}{$\Sigma_{\rm SFR}$}
\newcommand{\hi}{H{\sc i}}
\newcommand{\micron}{$\mu$m}
\newcommand{\taudepl}{$\tau_{\rm depl}$}
\newcommand{\msunpc}{${\rm M}_\odot\,{\rm pc}^{-2}$}

\title{The resolved star-formation relation in nearby active galactic nuclei}


\author{Viviana Casasola\inst{\ref{inst:vivi},\ref{inst:leslie}}
\and
Leslie Hunt\inst{\ref{inst:leslie}}
\and
Fran\c{c}oise Combes\inst{\ref{inst:francoise}}
\and
Santiago Garc\'{\i}a-Burillo\inst{\ref{inst:santi}}
}

\institute{INAF - Istituto di Radioastronomia \& Italian ALMA Regional Centre, Via Gobetti 101, 40129, Bologna, Italy
\label{inst:vivi}
\email{casasola@ira.inaf.it} 
\and
INAF - Osservatorio Astrofisico di Arcetri, Largo E. Fermi, 5, 50125, Firenze, Italy
\label{inst:leslie}
\and
Observatoire de Paris, LERMA (CNRS:UMR8112), 61 Av. de l'Observatoire, F-75014, Paris, France
\label{inst:francoise}
\and
Observatorio Astron\'omico Nacional (OAN)-Observatorio de Madrid,
Alfonso XII, 3, 28014-Madrid, Spain
\label{inst:santi}}

\date{Received ; accepted}

 
  \abstract
{}
{We present an analysis of the relation between the star formation rate (SFR) surface density (\sigmasfr) 
and mass surface density of molecular gas (\sigmahtwo), commonly referred to as the Kennicutt-Schmidt (K-S) relation,
on its intrinsic spatial scale, i.e. the size of giant molecular clouds ($\sim$10--150 pc), in the central, high-density regions 
of four nearby low-luminosity active galactic nuclei (AGN).
These are AGN extracted from the NUclei of GAlaxies (NUGA) survey.
This study investigates the correlations and slopes of the K-S relation, as a function of spatial
resolution and of the different $^{12}$CO emission lines used to trace \sigmahtwo, and tests its validity in the high-density central 
regions of spiral galaxies.
}
{We used interferometric IRAM $^{12}$CO(1--0) and $^{12}$CO(2--1) and SMA $^{12}$CO(3--2)  emission line maps 
to derive \sigmahtwo\ and  $HST$--H$\alpha$ images to estimate \sigmasfr.
}
{Each galaxy is characterized by a distinct molecular SF relation on spatial scales between 20 to 200~pc.
The K-S relations can be sublinear, but also superlinear, with slopes ranging from $\sim$0.5 to $\sim$1.3; 
slopes are generally superlinear on spatial scales $>$100~pc and sublinear on smaller scales.
Depletion times range from $\sim$1 and 2~Gyr, which is compatible with results for nearby normal galaxies.
These findings are valid independently of which transition -- $^{12}$CO(1--0), $^{12}$CO(2--1), or $^{12}$CO(3--2) --
is used to derive \sigmahtwo.
Because of either star-formation feedback, the lifetime of clouds, turbulent cascade, or magnetic fields,
the K-S relation might be expected to degrade on small spatial scales ($<$100~pc).
However, we find no clear evidence of this, even on scales as small as $\sim$20~pc,
and this might be because of the higher density of GMCs in galaxy centers that have to resist higher shear forces.
The proportionality between \sigmahtwo\ and \sigmasfr\ found
between 10 and 100~\msunpc\ is valid even at high densities, $\sim$10$^{3}$~\msunpc.
However, by adopting a common CO-to-H$_2$ conversion factor ($\alpha_{\rm CO}$), 
the central regions of the NUGA galaxies have higher \sigmasfr\ for a given gas column than those
expected from the models, with a behavior that lies between the mergers or high-redshift
starburst systems and the more quiescent 
star-forming galaxies, assuming that the first ones require a lower value of $\alpha_{\rm CO}$.
}
{}

\keywords{galaxies: ISM -- galaxies: spiral -- ISM: active -- ISM: molecules -- stars: formation 
}

   \maketitle
%

\section{Introduction}
\label{sec:intro}
The relationship between gas and star formation (SF) in galaxies plays a key role in galaxy evolution.  
It constrains how efficiently galaxies turn their gas into stars
and also serves as essential input to simulations and models
\citep[e.g.,][]{boissier03, tan99, springel03, krumholz05}.
Despite the importance of this relationship, the processes 
responsible for the conversion of gas into stars 
in various galactic environments are still poorly understood.

More than fifty years ago, \citet{schmidt59,schmidt63} suggested that the star formation rate (SFR) and the gas content 
in galaxies are related by
\begin{eqnarray}
\Sigma{\rm_{SFR}} = A\,\Sigma_{\rm gas}^{N}
\label{eq:schmidt}
\end{eqnarray}

\noindent
where \sigmasfr\ and $\Sigma_{\rm gas}$ are the SFR surface density and the  
gas surface density, respectively, $A$ is the normalization constant representing the efficiency of the processes
regulating gas-stars conversion, and $N_{\rm fit}$  the power relation index.
The gas can be atomic (\hi), molecular (H$_2$), or a combination of both (\hi+H$_2$).

Although the atomic phase of the interstellar medium (ISM) is directly traced by the \hi\ emission line at 21\,cm, 
indirect approaches are needed to estimate the distribution of H$_2$.
The molecular hydrogen indeed lacks a dipole moment and typical temperatures (15--25~K), in giant molecular clouds (GMCs)
are too low to excite quadrupole or vibrational transitions. 
For these reasons, the carbon monoxide (CO) emission lines are the most straightforward and reliable tracer of H$_2$ in galaxies. 
CO is relatively bright and its ability to trace the bulk distribution of H$_2$ has been
confirmed by comparisons with gamma rays \citep[][]{lebrun83,strong88} and dust emission \citep[][]{desert88,dame01}. 
Comparing the atomic gas density with the number of representative young stellar objects in the solar neighborhood, 
\citet{schmidt59} derived a power relation index $N \approx 2$ in equation (\ref{eq:schmidt}).  
Values of $N$ ranging approximately from 1.5 to 2 were further confirmed by \citet{guibert78}, using more precise data on
the radial and vertical distributions of the interstellar gas and a variety of young objects
as tracers of SFR.

Later, \citet{kennicutt98a,kennicutt98b} found that for disk-averaged surface densities, both 
normal star-forming and starburst galaxies follow equation (\ref{eq:schmidt}) with a power relation index of $N \approx 1.4$ for total 
(\hi+H$_2$) gas.
This correlation between gas and SFR surface density is commonly referred to as the Kennicutt-Schmidt relation (hereafter the K-S relation).
Such a relation is in principle consistent with large scale gravitational instability being the major driver 
of dense cloud formation \citep[][]{quirk72,madore77}.
Although spatially unresolved studies of \hi, CO, and SFR are useful for characterizing global disk properties, 
understanding the mechanisms behind the SF requires resolved measurements. 

It is now possible to study the K-S relation at 
sub-kpc scales, approaching the intrinsic scale of SF, i.e. the size of GMCs 
\citep[$\sim$10--150~pc, e.g.,][]{solomon87}.
Several factors have contributed, including 
the explosion in multiwavelength data for nearby galaxies: for example, UV-GALEX: Gil de Paz et al. 2007; 
infrared-\textit{Spitzer} and H$\alpha$-\textit{HST}: SINGS and LVL surveys, Kennicutt et al. (2003), Dale et al. (2009);  
far-infrared-Herschel: KINGFISH, Kennicutt et al. (2011); 
$^{12}$CO: BIMA SONG, Helfer et al. (2003); HERACLES-IRAM 30m, Leroy et al. (2009);
\hi: VLA THINGS survey, Walter et al. (2008).
Also technical improvements, especially the construction of millimeter interferometers,
now allow higher resolution imaging of $^{12}$CO in galaxies.

There is now substantial evidence that the molecular gas is
well-correlated with the SFR tracers over several orders of magnitude, 
but mostly for regions where H$_2$ makes up the majority of the neutral gas, 
$\Sigma_{\rm H2} \gtrsim \Sigma_{\rm HI}$ \citep[e.g.,][]{wong02,komugi05,kennicutt07,
thilker07,schuster07,bigiel08,leroy08, blanc09,onodera10,verley10,rahman11,momose13}.
This is because GMCs, the major reservoirs of molecular gas, are 
the sites of most SF in our Galaxy and other galaxies.
Their properties indeed set the initial conditions for protostellar collapse and may play a 
role in determining the stellar initial mass function \citep[IMF,][]{mckee07}.
Moreover, it has long been known that the spatial distribution of $^{12}$CO emission 
follows that of the stellar light and H$\alpha$ \citep[e.g.,][]{young82,scoville83,solomon83,tacconi90}.
The lack of a clear correlation between \sigmahi\ and \sigmasfr\ inside galaxy disks 
\citep[e.g.,][]{bigiel08} offers circumstantial evidence that SF remains coupled  to the molecular,
rather than total gas surface density ($\Sigma_{\rm HI+H2}$) even where \hi\ makes up most of the ISM.
\citet{bigiel08} find that \sigmahi\ saturates at surface density of $\approx$9~M$_{\odot}$~pc$^{-2}$,
and gas in excess of this value is in the molecular phase of spirals and \hi-dominated galaxies. 

A crucial parameter in the study of the resolved K-S relation is the choice of the tracer 
for H$_2$ and SFR.
Most of the studies for deriving the H$_2$ surface density to
study the molecular SF relation are based on the 
$^{12}$CO(1--0) \citep[e.g.,][]{rahman11,saintonge11,momose13} or $^{12}$CO(2--1) emission lines
\citep[e.g.,][]{bigiel08,leroy08}, and more rarely on the $^{12}$CO(3--2) transition \citep[e.g.,][]{komugi07,wilson12}. 
Also the dependence of the SFR on dense molecular gas mass has been explored with molecules such as HCN and HCO$^{+}$ 
\citep[e.g.,][]{gao04,wu05,santi12}.

Over the past thirty years, extensive efforts have been made to 
derive plausible SFRs for external galaxies \citep[see][for reviews]{kennicutt98a,calzetti10}.
Optical SFR tracers, such as H$\alpha$, often suffer from dust extinction, which can change dramatically from
location to location.
\citet{calzetti07} find $A_V \sim 2.2$~mag in typical extragalactic H{\sc ii} regions and dense star-forming
regions can be completely obscured with $A_V$ that can reach $\sim$6~mag \citep[e.g.,][]{scoville01}.   
In addition, the $A_V$ value is not always known for a given galaxy.
However, IR space facilities (\textit{Spitzer} and \textit{Herschel}) and UV (GALEX) have 
made it possible to also image the details of dust-obscured SF. 

Different SFR tracers probe different time scales and hence the SF history of any particular galaxy.
H$\alpha$ emission traces ionized gas by massive (M~$> 10$~M$_{\odot}$) stars on a timescale
of $<$~20~Myr.
The far-UV (FUV) luminosity corresponds to relatively older ($< 100$~Myr), less massive ($\gtrsim 5$~M$_{\odot}$)   
stellar populations.
The MIR 24\,\micron\ emission mostly traces reprocessed radiation of newborn (few Myr) OB stellar associations 
embedded within the parental molecular clouds.
Although star clusters emerge from their natal clouds in less than 1~Myr, they remain associated with it on 
a much longer time scale, $\sim$10--30~Myr, which is the time scale associated with the  24\,\micron\ emission 
as a SFR tracer.
\citet{kennicutt07} and \citet{calzetti07} have justified the feasibility of correcting the number of ionizing 
photons, as traced by the H$\alpha$ recombination line, for the effects of the dust extinction by adding a weighted 
component from the \textit{Spitzer} MIPS 24\,\micron\ luminosity in individual star-forming regions.
\citet{leroy08} and \citet{bigiel08} have proposed another composite SFR tracer that corrects dust attenuation of far-UV
surface brightness using, also in this case, the \textit{Spitzer} 24\,\micron\ emission but with different weights.
These composite SFR tracers,  H$\alpha$+24\,\micron\ and FUV+24\,\micron, have been extensively used
for a large number of nearby galaxies (see references above). 

Spatially resolved K-S relation studies published in the past decade 
\citep[e.g.,][]{kennicutt07,bigiel08,leroy08,blanc09,bigiel10,eales10,verley10,bigiel11,liu11,rahman11,rahman12,schruba11,ford13,viane14} 
found a wide spread in the value of the power relation index ($N \approx 0.6 - 3$) both within and among galaxies.
For a comprehensive review of the most recent SF relation studies on sub-kpc scales, we refer to \citet{kennicutt12}.
The wide range of the value of $N$ may be intrinsic, suggesting that different SF `laws' exist, and may contain valuable 
astrophysical information.
Alternatively, it may be partially or entirely due to the adopted physical scale \citep[e.g.,][]{schruba10,calzetti12},
the choice of the molecular gas \citep[e.g.,][]{krumholz07,narayanan08} and SFR tracer, the type of galaxy under investigation, 
data sampling, and the fitting method used \citep[e.g.,][]{blanc09,shetty13}.
In particular, different SFR tracers and spatial scales effectively sample different timescales, so that
a galaxy's SF history can play a role in determining the results of the measurements. 
It is also possible that these differences correspond to a spectrum of physical mechanisms present in a wide range of
environments. 
High shear in galactic bars, harassment in a dense galaxy cluster, and galaxy mergers and interactions 
have the potential to either dampen or enhance the SF process \citep[e.g.,][]{vivi04,zhou11,lanz13}.
All these environmental processes at work, on both galactic and extragalactic
scales, suggest therefore that there is no universal SF relation in the Local Universe. 

In addition to the slope of the empirical power relation relation between $\Sigma_{\rm gas}$ and \sigmasfr, 
the other crucial parameter in SF studies is the molecular gas depletion time, defined as the time needed 
for the present SFR to consume the existing molecular gas reservoir, 
$\tau_{\rm depl}\,=\,\Sigma_{\rm gas}/\Sigma_{\rm SFR}$, i.e. the inverse of the star formation efficiency (SFE).
One interpretation of the linear slope for the K-S relation is an approximately constant  \taudepl, 
with an average \taudepl\ of about 2~Gyr in normal spirals \citep[e.g.,][]{leroy13}.
As for the power relation index, several factors complicate the interpretation of \taudepl.
Moreover, while a linear slope (and constant \taudepl) describes the global average scaling relation
in the star-forming galaxies well, individual galaxies deviate from a single \taudepl\
\citep[e.g.,][]{saintonge11,leroy13}.

The objective of this paper is to explore the molecular SF relation in the inner ($\sim$1~kpc), high-density regions 
of four nearby ($D \lesssim 20 $ Mpc), low-luminosity active galactic nuclei (LLAGN) on the spatial scale of $\sim$20--200~pc 
through a pixel-by-pixel analysis of the available maps. 
We do not discuss the correlation of SFR with the atomic component of the gas, since 
the gas phase is predominantly molecular in the central regions we study.
These galaxies were originally part of the NUclei of GAlaxies (NUGA) survey carried out
at the Plateau de Bure Interferometer \citep[PdBI,][]{santi03}.
The good spatial resolution offered by $^{12}$CO NUGA observations enables us
to probe GMC spatial scales around LLAGN, and better understand the SF process.
Here, we investigate the correlations and slopes of the SF relation as a function of spatial
resolution and of various $^{12}$CO transitions (1--0, 2--1, 3--2) for deriving \sigmahtwo, and 
the validity of the K-S relation in regions with high molecular gas surface densities. 

The NUGA sample is ideal for such a study for several reasons: 
The proximity of the galaxies combined with the good spatial resolution and sensitivity afforded by PdBI 
and SMA and the high-resolution SFR tracers from the \textit{Hubble Space Telescope} (\textit{HST}) let us
probe the K-S relation on fine spatial scales that up to now have
only been examined in Local Group galaxies.
Moreover, the physical conditions in the NUGA LLAGN are more extreme than those found in typical spiral disks.
We can thus assess how well the warm dense gas at high column densities 
in AGN circumnuclear regions can form stars.
These findings set the stage for what will be possible with ALMA data. 

This paper is organized as follows. 
We describe the sample selection in Section \ref{sec:sample}, and the data we used, 
their treatment, the procedure for deriving \sigmahtwo\ and \sigmasfr\ maps from the original images, 
and the applied fitting method in Sect.~\ref{sec:data}.
In Sects.~\ref{sec:sf-law} and  \ref{sec:sf-sample} we show the results on the observed relationships between \sigmahtwo\ and \sigmasfr\ 
for individual galaxies and the whole sample, respectively. 
In Sect.~\ref{sec:caveats} we stress the caveats and uncertainties associated with the present study, 
and finally in Sect.~\ref{sec:conclusion} we summarize our findings and give our conclusions.

\section{The sample}
\label{sec:sample}
The sample presented in this paper consists of four LLAGN selected from the NUGA
survey.
The NUGA project is an IRAM Plateau de Bure Interferometer (PdBI) and 30 m single-dish survey 
of nearby LLAGN with the aims (i) of mapping, at the angular resolution of $\sim$0.5-2\arcsec\ and sensitivity 
of $\sim$2-4~mJy\,beam$^{-1}$, the distribution and dynamics of the molecular gas through 
the two lowest emission lines of the $^{12}$CO in the inner kpc of the galaxies, and (ii) of studying the different 
mechanisms for gas fueling of LLAGN. 
Each galaxy of the core NUGA sample (12) has been studied on a case-by-case basis.

In this paper, we present the results of the study of the molecular gas spatially-resolved SF relation 
in the following NUGA galaxies:  NGC~3627, NGC~4569, NGC~4579, and NGC~4826. 
This NUGA subsample offers heterogeneity in terms of nuclear activity, distance, detection or not 
of gas inflow, gas morphology, and surrounding environment.
Since the sample only consists of four galaxies, the wide range of galaxy properties cannot be used to infer statistical conclusions. 
It is worthwhile, however, looking at the different conditions where the SF relation is studied in the selected NUGA subsample. 
These properties are collected in Table~\ref{tab:sample}.
In this table, Col. (1) indicates the galaxy name, Cols. (2) and (3) the coordinates
of the galaxy dynamical center derived from NUGA IRAM $^{12}$CO observations (see later),
Col. (4) the morphological type from the Third Reference Catalog of Bright Galaxies \citep[RC3,][]{devaucouleurs91},
Col. (5) the nuclear activity, Col. (6) the distance ($D$), Col. (7) the inclination ($i$), Col. (8)
the position angle (PA), Col. (9) the prevalent molecular gas morphology as detected from NUGA observations, 
Col. (10) the identified molecular gas motion (i.e., inflow), Col. (11) the surrounding environment, 
and Col. (12) the NUGA references.

\begin{sidewaystable*}
\caption[]{Properties of the galaxy sample.}
\begin{center}
\begin{tabular}{cccccccccccc}
\hline
\hline
Galaxy & $\alpha_{\rm J2000}$ & $\delta_{\rm J2000}$ & 
RC3 type & Nuclear act. & $D$ & $i$ & PA & Gas morph. $^{\mathrm{(a)}}$ & Gas mot. & Envir. & NUGA Ref. $^{\mathrm{(b)}}$\\
& [$^{\rm h}$ $^{\rm m}$ $^{\rm s}$]  & [$^{\circ}$ $^{\prime}$ $^{\prime\prime}$]  & &  & [Mpc] & [deg]  & [deg]\\
(1) & (2) & (3) & (4) & (5) & (6) & (7) & (8) & (9) & (10) & (11) & (12) \\
\hline
NGC 3627 & 11 20 15.02 & 12 59 29.5
& SAB(s)b & LINER/Seyfert 2 & 10.2 (1\arcsec=49\,pc) & 61 & 178 & Bar & Inflow & Interacting & (1)\\
NGC 4569 & 12 36 49.80 & 13 09 46.3
& SAB(rs)ab & Transition 2 &  16.8 (1\arcsec=81\,pc) & 62 & 15  & Bar & Inflow & Isolated & (2,3)\\
NGC 4579 & 12 37 43.52 & 11 49 05.5 
& SAB(rs)b & LINER/Sy1.9 & 19.8 (1\arcsec=96\,pc) & 36 & 95 & Spiral arms & Inflow & Isolated & (4,5)\\
NGC 4826 & 12 56 43.64 & 21 40 59.3 
& (R)SA(rs)ab & LINER & 5.4 (1\arcsec=26\,pc) & 54 & 112 & Disk & -- & Isolated & (3,4,6)\\
\hline
\hline
\end{tabular}
\label{tab:sample}
\end{center}
\tablefoot{
\tablefoottext{a}{Molecular gas morphology as revealed from NUGA IRAM $^{12}$CO(1--0) observations. 
For $^{12}$CO(2--1) and $^{12}$CO(3--2) morphologies we refer to NUGA papers.}
\tablefoottext{b}{(1) \citet{vivi11};
(2) \citet{boone07};
(3) \citet{boone11};
(4) \citet{santi05};
(5) \citet{santi09};
(6) \citet{santi03}.}
}
\end{sidewaystable*}

\section{The data, derived parameters, and fitting method}
\label{sec:data}
This study is based on measurements of the molecular gas and SFR surface densities,
\sigmahtwo\ and \sigmasfr, and relies on
the existence of multiwavelength datasets.
We used $^{12}$CO(1--0), $^{12}$CO(2--1), and $^{12}$CO(3--2) line intensity maps to 
derive the surface densities of the molecular gas, and $HST$ H$\alpha$ (6563~$\AA$) emission 
images to estimate the surface densities of SF. 
In this section we describe these datasets. 

\subsection{IRAM $^{12}$CO(1--0) and $^{12}$CO(2--1) observations}
\label{sec:iram-data}
The $^{12}$CO(1--0) and $^{12}$CO(2--1) line intensity maps are part of the NUGA survey.
NUGA observations have been carried out with six antennas of the PdBI 
in the ABCD configuration of the array and with the IRAM 30m single-dish telescope. 
$^{12}$CO images were reconstructed using the standard 
IRAM/GILDAS\footnote{http://www.iram.fr/IRAMFR/GILDAS/} software \citep{guilloteau},
following the prescriptions described in \citet{santi03}, and
restored with Gaussian beams.
The beam sizes are typically $\la$2\arcsec\ for $^{12}$CO(1--0) and 
$\la$1\arcsec\ for $^{12}$CO(2--1).
We used natural and uniform weightings to generate $^{12}$CO(1--0) 
and $^{12}$CO(2--1) maps, respectively. 
This allowed us to maximize the flux recovered in $^{12}$CO(1--0) and optimize the spatial 
resolution in $^{12}$CO(2--1). 
For NGC~3627 and NGC~4579, 30m observations were used to compute the short spacings and complete the 
interferometric measurements, whereas for NGC~4569 and NGC~4826,
we used $^{12}$CO maps obtained only from PdBI observations. 

For each NUGA galaxy we derived the position of the AGN by assuming that this
coincides with the dynamical center of the galaxy (see Table \ref{tab:sample}).
This choice maximizes the symmetry of the global velocity field derived from the two $^{12}$CO transitions.
Details on the determination of the AGN position/dynamical center
are discussed on a case-by-case NUGA papers.
Properties of IRAM $^{12}$CO maps are collected in Table~\ref{tab:info-co}, 
and described in detail in Sect.~\ref{sec:h2}.

\subsection{SMA $^{12}$CO(3--2) observations}
\label{sec:sma-data}
For NGC~4569 and NGC~4826, we also have $^{12}$CO(3--2) line intensity maps obtained 
with the Submillimeter Array (SMA) in its compact configuration with seven working antennas.
$^{12}$CO(3--2) single-dish observations carried out with the 15~m James Clerk Maxwell Telescope (JCMT)
and published by \citet{wilson09} for NGC~4569 and \citet{israel09} for NGC~4826
were used to compute and add the missing short spacings to the SMA data.
The beam sizes of $^{12}$CO(3--2) maps are comparable to 
those of the IRAM $^{12}$CO(1--0) maps, i.e. $\sim$2\arcsec\ (see Table~\ref{tab:info-co}).
Details on these observations are contained in \citet{boone11}.

\begin{sidewaystable*}
\caption[]{Basic information on $^{12}$CO NUGA dataset.}
\begin{center}
\begin{tabular}{ccccccc}
\hline
\hline
Galaxy & IRAM instrument & $^{12}$CO line & Beam & 1$\sigma$ & $F{\rm_{10}}$, $F{\rm_{21}}$, $F{\rm_{32}}$ & $R_{21}$, $R_{32}$ \\ 
& & & [$^{\prime\prime}$ $\times$ $^{\prime\prime}$, $^\circ$] & [Jy\,beam$^{-1}$\,km\,s$^{-1}$] & [K\,(Jy\,beam$^{-1}$)$^{-1}$] & \\ 
        (1) & (2) & (3) & (4) & (5) & (6) & (7) \\
\hline
NGC 3627 & PdBI+30m & (1--0) & 2.05 $\times$ 1.26, 30 & 0.16 & 35.55 & --    \\
         & PdBI+30m & (2--1) & 0.86 $\times$ 0.58, 23 & 0.30 & 46.32 & 0.60  \\
\hline
NGC 4569 & PdBI     & (1--0) & 2.33 $\times$ 1.46, 27 & 0.25 & 27.00 & --    \\
         & PdBI     & (2--1) & 1.17 $\times$ 0.65, 198 & 0.35 & 30.24 & 0.63  \\
         & SMA+JCMT & (3--2) & 2.60 $\times$ 2.06, $-36$ & 6.19 & 1.92  & 0.23  \\
\hline
NGC 4579 & PdBI+30m & (1--0) & 2.03 $\times$ 1.28, 26.03 & 0.16 & 35.34 & --    \\      
         & PdBI+30m & (2--1) & 0.98 $\times$ 0.60,  $-155.48$& 0.32 & 38.94 & 0.90  \\      
\hline
NGC 4826 & PdBI     & (1--0) & 2.53 $\times$ 1.80 & 0.25 & 20.17 & --    \\      
         & PdBI     & (2--1) & 0.77 $\times$ 0.55 & 0.36 & 54.21 & 0.67  \\
         & SMA+JCMT & (3--2) & 2.58 $\times$ 1.94 & 4.44 & 2.05  & 0.38  \\
\hline
\hline
\end{tabular}
\label{tab:info-co}
\end{center}
\end{sidewaystable*}

\subsection{$HST$ H$\alpha$ emission-line images}
\label{sec:Ha-data}
We retrieved H$\alpha$ emission-line images from the Hubble Legacy Archive (HLA) that makes 
\textit{HST} WFPC2 observations of our galaxy sample available.
In these images, the H$\alpha$ emission line was observed through the narrow-band filters F656N or F658N, 
and the underlying continuum through F547N, F555W, and/or F814W (equivalent 
to narrow-band $V$, $V$, and $I$, respectively).
The $HST$ images have a pixel size of 0\farcs05 and a spatial resolution of 
0\farcs1.

The available maps are emission-line-only H$\alpha$+[N{\sc ii}] images 
containing emission from both H$\alpha$ at 6563~$\AA$
and [N{\sc ii}]$\lambda\lambda$~6548, 6584~$\AA$.
We removed the [N{\sc ii}] contamination within the filter bandpass
using average [N {\sc ii}]/H$\alpha$ values available in literature.
Then, we corrected the H$\alpha$ maps both for Galactic foreground extinction
and internal extinction.
We assumed the values of the $B$-band Galactic foreground extinction $A_{B}$(Gal) available from the literature
(see Table~\ref{tab:info-halpha}) and used the interstellar extinction curve by \citet{cardelli89}.
For three galaxies of our sample (NGC~3627, NGC~4569, and NGC~4826), internal extinction corrections 
were extracted from \citet{calzetti07} using Pa$\alpha$ images as a yardstick for calibrating the MIR emission.
The hydrogen emission lines trace the number of ionizing photons, and the Pa$\alpha$ line (at 1.8756~\micron)
is only modestly affected by dust extinction.
Because of its relative insensitivity to dust extinction \citep[less than a factor 2 of correction for typical extinction
in nearby galaxies, $A_V \lesssim5$~mag,][]{calzetti07}, Pa$\alpha$ is a nearly unbiased tracer of
the current SFR \citep[][]{kennicutt98b}.
Among the various SFR calibrations, the linear combination of the observed H$\alpha$ and the 
24\,\micron\ emission is the one most tightly correlated with the extinction-corrected Pa$\alpha$ emission 
\citep[][]{calzetti07}.
The most straightforward interpretation of this trend is that the 24\,\micron\ 
emission traces the dust-obscured SF, while the observed H$\alpha$ emission traces the unobscured one \citep{kennicutt07}.   
Thus, the combination of the two recovers all the SF in a given region.
This interpretation is confirmed by models 
(e.g., STARBURST99, 2005 update, Leitherer et al. 2005; Draine \& Li 2007), which also suggest that the trend is 
relatively independent of the characteristics of the underlying star-forming population.
This implies that H$\alpha$/Pa$\alpha$ ratio and the internal extinction derived from it, for instance $A_{V}$(int) 
in the $V$-band as performed by \citet{calzetti07}, are good tracers of the internal extinction of a galaxy.   

For NGC~4579, rather than the H$\alpha$/Pa$\alpha$ ratio,
we used the internal extinction derived from the Balmer decrement by \citet{ho99}.
One of the most reliable techniques for estimating interstellar extinction is indeed to measure the 
flux ratio of two nebular Balmer emission lines such as H$\alpha$/H$\beta$ (i.e., the Balmer decrement).
The determination of dust extinction from the Balmer decrement has been shown to be a very successful 
technique in the Local Universe since the first statistical work by \citet{kennicutt92}. 
These results have been improved upon by the large amount of optical spectra provided by the Sloan Digital
Sky Survey (SDSS), which were analyzed in this context by \citet{brinchmann04} and \citet{garn10}.
The main limitation to applying the internal extinction correction derived from Balmer decrement is to neglect 
a possible completely obscured SFR component, hence to underestimate the total SFR.
In any case, NGC~4579 has a $A_{V}$(int) value similar to those of the other galaxies, 
as shown in Table~\ref{tab:info-halpha}.
This table collects the properties of the original H$\alpha$+[N{\sc ii}] images and the values 
of the parameters used to obtain final H$\alpha$ maps.
In Table~\ref{tab:info-halpha}, Col. (1) indicates the galaxy name, Col. (2) the $HST$ instrument, 
Col. (3) the 1$\sigma$ noise of the background subtracted H$\alpha$+[N{\sc ii}] images, 
Col. (4) the adopted [N{\sc ii}]/H$\alpha$ ratio, 
Col. (5) the $B$-band Galactic foreground extinction [$A_{B}$(Gal)], and 
Col. (6) the $V$-band internal extinction [$A_{V}$(int)].

\begin{table*}
\caption[]{Basic information on the $HST$-H$\alpha$ dataset.}
\begin{center}
\begin{tabular}{cccccccc}
\hline
\hline
Galaxy & $HST$ instrument & 1$\sigma$                           & [N{\sc ii}]/H$\alpha$ & $A_{B}$(Gal)  & $A_{V}$(int)   \\
            &                                & [erg\,s$^{-1}$\,cm$^{-2}$]   &                                     & [mag]              & [mag]               \\
(1)      & (2)          & (3)                   & (4)       & (5) & (6)\\
\hline
NGC 3627 & WFPC2(F658N) & $5.43\times10^{-19}$ & 0.54 $^{\mathrm{(a)}}$ & 0.08 $^{\mathrm{(a)}}$ & 1.94 $^{\mathrm{(b)}}$   \\
NGC 4569 & WFPC2(F656N) & $1.90\times10^{-18}$ & 0.50 $^{\mathrm{(b)}}$ & 0.20 $^{\mathrm{(c)}}$ & 1.05 $^{\mathrm{(b)}}$  \\
NGC 4579 & WFPC2(F658N) & $9.49\times10^{-18}$ & 0.62 $^{\mathrm{(d)}}$ & 0.18 $^{\mathrm{(c)}}$ & 2.10 $^{\mathrm{(e)}}$   \\
NGC 4826 & WFPC2(F656N) & $6.19\times10^{-18}$ & 0.51 $^{\mathrm{(a)}}$ & 0.16 $^{\mathrm{(a)}}$ & 1.97 $^{\mathrm{(b)}}$  \\
\hline
\hline
\end{tabular}
\label{tab:info-halpha}
\end{center}
\tablefoot{
\tablefoottext{a}{\citet{kennicutt08}. These values
are an average of the values from the maps of \citet{burstein82} and \citet{schlegel98}.}
\tablefoottext{b}{\citet{calzetti07}.}
\tablefoottext{c}{From NED.}
\tablefoottext{d}{\citet{kennicutt09}.}
\tablefoottext{e}{\citet{ho99}.}
}
\end{table*}

\subsection{Image treatment}
\label{sec:treatment}
With the images described above, we constructed maps of
\sigmahtwo\ and \sigmasfr\ to perform a pixel-by-pixel analysis of 
the spatially resolved K-S relation. 
The procedures for obtaining the final maps are described later in Sections \ref{sec:h2} and \ref{sec:sfr}.
Since we compared images with different properties and wide spreads in resolution, the first task 
was to convert them to a common alignment and resolution.

All the original images ($^{12}$CO(1--0), $^{12}$CO(2--1), $^{12}$CO(3--2), and H$\alpha$) 
were centered on the dynamical centers of the galaxies derived from NUGA IRAM $^{12}$CO 
observations (see Table \ref{tab:sample}).
The H$\alpha$ maps were convolved to the resolution of the $^{12}$CO (1--0, 2--1, and 3--2) maps 
with a Gaussian beam (on the sky); i.e., we do not account for the inclination of the galaxy.
Then, all maps have been resampled to a pixel size equal to the adopted $^{12}$CO resolution
\citep[see, for instance,][for a similar treatment of images]{vutisalchavakul14}.
Since the spatial resolution and pixel size are equivalent, the pixels can be
considered as roughly statistically independent, and there should be little correlation among them.
Each pixel was thus treated as a single data point.
Finally, we also convolved all maps to a common 200~pc resolution
to be able to distinguish between the effects of different $J$ transitions and
spatial resolution.
These procedures were performed by using IRAM/GILDAS and IRAF\footnote{IRAF is the Image Reduction and Analysis Facility. 
IRAF is written and supported by the National Optical Astronomy Observatories (NOAO) in Tucson, 
Arizona. NOAO is operated by the Association of Universities for Research in Astronomy (AURA), Inc. 
under cooperative agreement with the National Science Foundation} softwares.

We thus probed the gas and SFR surface densities in individual galaxies on physical scales ranging 
from 17\,pc to 190\,pc, as well as with a common resolution of 200\,pc. 
Such scales are smaller than those previously scrutinized
outside the Local Group \citep[e.g.,][]{bigiel08,leroy08,rahman11} 
and are superior in this respect 
to other kinds of analyses, such as azimuthally averaged radial profiles 
\citep[e.g.,][]{kennicutt89,bigiel08,rahman11}
or the aperture analysis encompassing star-forming regions and
centering on H$\alpha$ emission peaks \citep[e.g.,][]{kennicutt07,blanc09}.

\subsection{Molecular gas surface density maps}
\label{sec:h2}
The availability of $^{12}$CO(1--0), $^{12}$CO(2--1), and $^{12}$CO(3--2) data offers \sigmahtwo\ maps 
at different resolutions and the possibility to compare the three lowest $^{12}$CO transitions in terms of 
the K-S relation.
We derived \sigmahtwo\ maps from $^{12}$CO(1--0), $^{12}$CO(2--1), and $^{12}$CO(3--2) 
integrated intensity maps ($S\rm_{CO(1-0)}$, $S\rm_{CO(2-1)}$, $S\rm_{CO(3-2)}$) by adopting a constant value for the 
$X_{\rm CO}$ conversion factor, $ X_{\rm CO} = 2.2 \times 10^{20}$ cm$^{-2}$ (K km s$^{-1}$)$^{-1}$ \citep[][]{solomon91}
that corresponds to $\alpha_{CO} = 3.5$~$M_{\odot}$~pc$^{-2}$~(K~km~s$^{-1}$)$^{-1}$ 
\citep[e.g.,][]{narayanan12,bolatto13}.
For $^{12}$CO(1--0) emission, the conversion to \sigmahtwo\ is 
\begin{eqnarray}
\Sigma\rm_{H_{2}} &=& 3.5\,S{\rm_{CO(1-0)}}\,F{\rm_{10}}\,{\rm cos}\,i
\label{s10}
\end{eqnarray}

\noindent
where \sigmahtwo\ is in units of M$_{\odot}$\,pc$^{-2}$, 
$S\rm_{CO(1-0)}$ in Jy\,beam$^{-1}$\,km\,s$^{-1}$,
$F{\rm_{10}}$ 
is the conversion factor from flux density to 
brightness temperature for the $^{12}$CO(1--0) line in K\,(Jy\,beam$^{-1}$)$^{-1}$, 
and $i$ is the galaxy inclination.
For $^{12}$CO(2--1) emission, the conversion to $\Sigma _{{\rm H}_2}$ is derived from equation (\ref{s10})
by defining $R_{21}$ as the $^{12}$CO(2--1)/$^{12}$CO(1--0) line ratio in temperature units: 
\begin{eqnarray}
\Sigma\rm_{H_{2}} &=& 3.5\,\frac{S{\rm_{CO(2-1)}}}{R{\rm_{21}}}\,F{\rm_{21}}\,{\rm cos}\,i
\label{s21}
\end{eqnarray}

\noindent
where $F{\rm_{21}}$ is the conversion factor from flux density to 
brightness temperature for the $^{12}$CO(2--1) line.
Consistently with equation (\ref{s21}), for $^{12}$CO(3--2) emission the conversion to 
$\Sigma _{{\rm H}_2}$ is derived by defining $R_{32}$ as the $^{12}$CO(3--2)/$^{12}$CO(1--0) 
line ratio: 
\begin{eqnarray}
\Sigma\rm_{H_{2}} &=& 3.5\,\frac{S{\rm_{CO(3-2)}}}{R{\rm_{32}}}\,F{\rm_{32}}\,{\rm cos}\,i
\label{s32}
\end{eqnarray}

\noindent
where $F{\rm_{32}}$ is the conversion factor from flux density to 
brightness temperature for the $^{12}$CO(3--2) line.
Both $R_{21}$ and $R_{32}$ used in the equations (\ref{s21}) and (\ref{s32}), respectively,
have been measured for each NUGA galaxy; we used the mean
value of these measured ratios to convert all transitions to CO(1--0).
Equations (\ref{s10}), (\ref{s21}), and (\ref{s32}) define \textit{\emph{hydrogen}} surface densities; i.e., 
they do not include any contribution from helium.
To scale our quoted surface densities to account for helium, they should be multiplied by a factor $\sim$1.36.

The values needed to apply equations (\ref{s10}), (\ref{s21}), and (\ref{s32}) are listed in Tables 
\ref{tab:sample} and \ref{tab:info-co}.
In Table~\ref{tab:info-co}, Col. (1) gives the galaxy name, Col. (2) the instruments used for observations, 
Col. (3) the observed $^{12}$CO transition, Col. (4) the Gaussian beam FWHM, Col. (5) the 1$\sigma$ noise of the image,
Col. (6) the conversion factor from flux density to brightness temperature for the three $^{12}$CO transitions, and
Col. (7) the $R_{21}$ and $R_{32}$ line ratios.

\subsection{Star formation rate surface density maps}
\label{sec:sfr}
Several SFR calibrations using different tracers, 
based on a variety of galaxy and/or region samples and stellar IMFs, have been published 
\citep[e.g.,][]{wu05,calzetti07,zhu08,rieke09}.
As stated above, the H$\alpha$ recombination emission line provides a nearly 
instantaneous measure of the SFR independently of the previous SF history. 
Among the hydrogen recombination lines, H$\alpha$  is the most widely used as SFR tracer
because of its higher intensity and lower sensitivity to dust
attenuation than bluer nebular lines (e.g., Ly$\alpha$, Ly$\beta$, H$\beta$).

To derive SFR maps we used the conversion between SFR and dust extinction-corrected H$\alpha$ flux density derived 
by \citet{calzetti07}:
\begin{eqnarray}
\begin{split}
\Sigma{\rm_{SFR}} & = 5.3 \times 10^{-42}\,S({\rm H\alpha)_{corr}}\,{\rm cos}\,i & 
\label{sfr-halpha}
\end{split}
\end{eqnarray}

\noindent
where \sigmasfr\ is in units of M$_{\odot}$\,yr$^{-1}$\,kpc$^{-2}$, and 
$S$(H$\alpha$)$_{\rm corr}$ is the dust extinction-corrected H$\alpha$ in erg~s$^{-1}$~kpc$^{-2}$.
This calibration has been derived assuming the stellar IMF of STARBURST99, which consists of two power laws with slope $-1.3$ in the range 
$0.1 - 0.5$~M$_{\odot}$ and slope $-2.3$ in the range $0.5 - 120$~M$_{\odot}$.
 \citep[For details on the adopted models of stellar populations see Appendix A2 in][]{calzetti07}

We have also checked that the H$\alpha$ emission is uncontaminated by emission from the AGN 
by comparing different (albeit lower resolution) estimators of SFR.
There is virtually no contamination from AGN-excited H$\alpha$ emission
outside the central pixel for NGC~3627 and NGC~4826.
Although NGC~4826 has no X-ray source and NGC~3627 is only weakly detected
\citep{hernandez13}, NGC~4569 and NGC~4579 both have weak nuclear X-ray sources 
\citep{dudik05}.
In NGC~4579 potential AGN contamination may be more of a problem because of the 
broad H$\alpha$ emission \citep{ho97}.
Nevertheless, the SFR inferred from broad H$\alpha$ is $\sim$0.02\,M$_\odot$~yr$^{-1}$,
while the total SFR for this galaxy is $\sim$2.2\,M$_\odot$~yr$^{-1}$ (see Table \ref{tab:ssfr}),
more than 100 times greater.
 We therefore conclude that SF processes are dominating the AGN
in our sample galaxies.
For safety, in NGC~4569 and NGC~4579 we masked the central $2 \times 2$ pixels 
since there was the possibility that the strong H$\alpha$ emission there is due to the AGN.

\subsection{Fitting method}
\label{sec:fit}
With the images described above, we constructed maps of
\sigmahtwo\ and \sigmasfr\ to perform a pixel-by-pixel analysis of 
the spatially resolved K-S relation. 
We used data above 3$\sigma$ significance both in \sigmahtwo\ and \sigmasfr\ maps.

We fit the data in logarithmic space:
\begin{eqnarray}
{\rm log}(\Sigma{\rm_{SFR}}) = A_{\rm fit} + N_{\rm fit}\,\times\, {\rm log}(\Sigma_{\rm H_2})
\label{log}
\end{eqnarray}

\noindent
where $A_{\rm fit}$ is the intercept and $N_{\rm fit}$  the index of the K-S relation.
We use the ordinary least squares (OLS) linear bisector method 
\citep[][]{isobe90,feigelson92}, adopted in several SF relation studies \citep[e.g.,][]{bigiel08,schruba11,momose13}.

For statistical completeness, we also give results derived from a robust regression fitting 
method \citep[][]{li85,fox97}, an alternative to least squares regression, which provides increased uncertainties 
of slopes and intercepts with respect to those given by the OLS bisector method.
These fits are implemented in the public-domain statistical software package $R$ \citep{ihaka96}.
However, in the following, we present and discuss results emerging from the OLS bisector method 
for reasons described later in Sect.~\ref{sec:fitting-choice}.   

\begin{figure*}[!htp]
\centering
\includegraphics[width=0.6\textwidth]{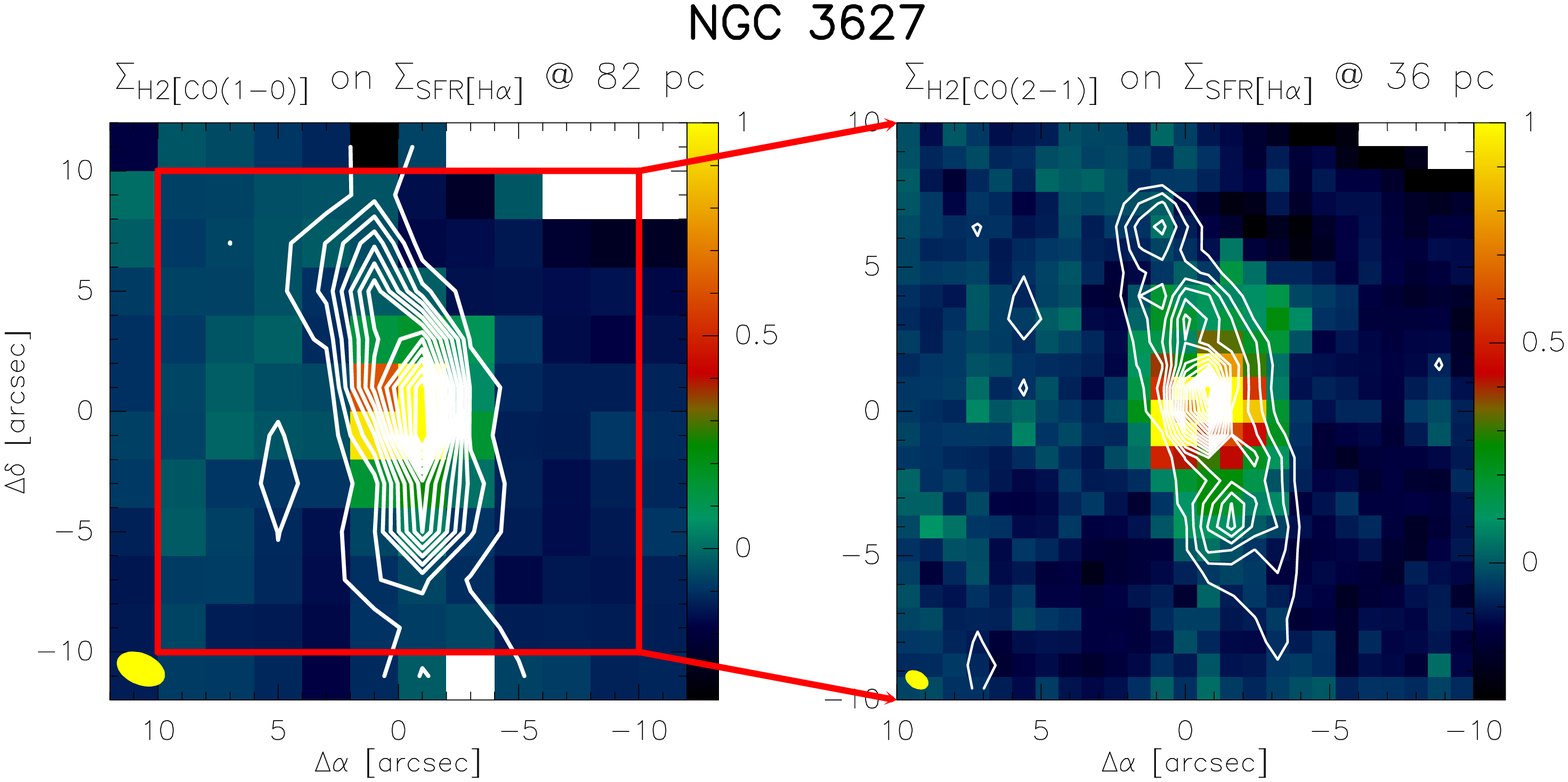}
\includegraphics[width=1\textwidth]{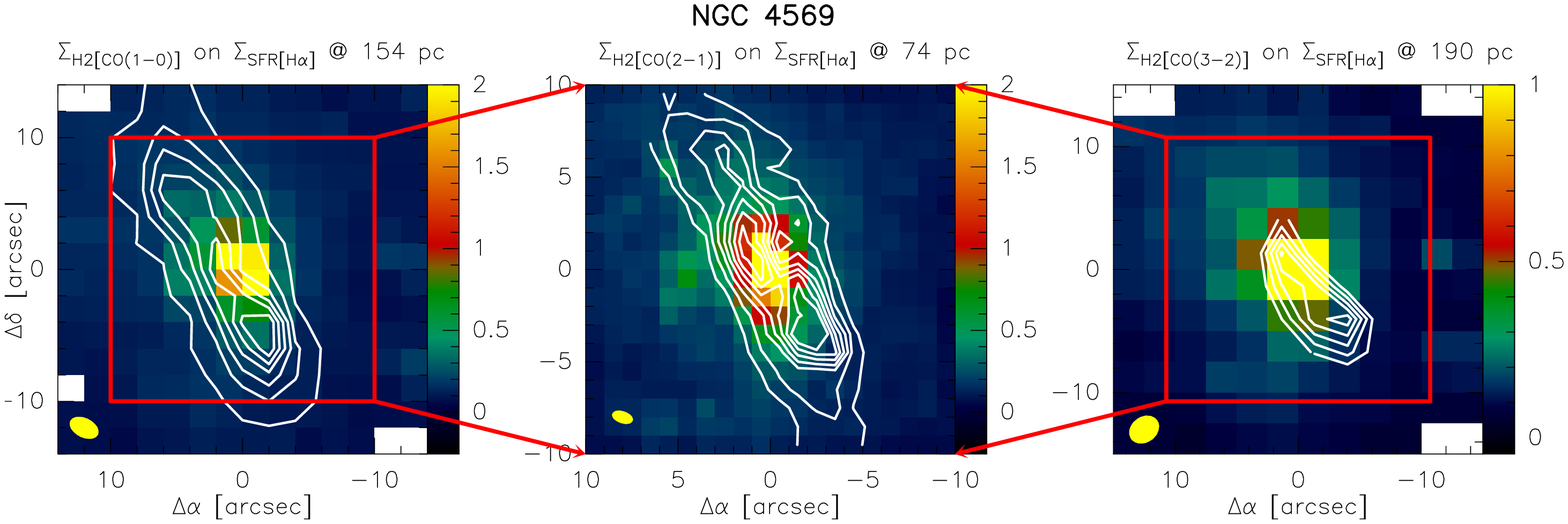}
\includegraphics[width=0.6\textwidth]{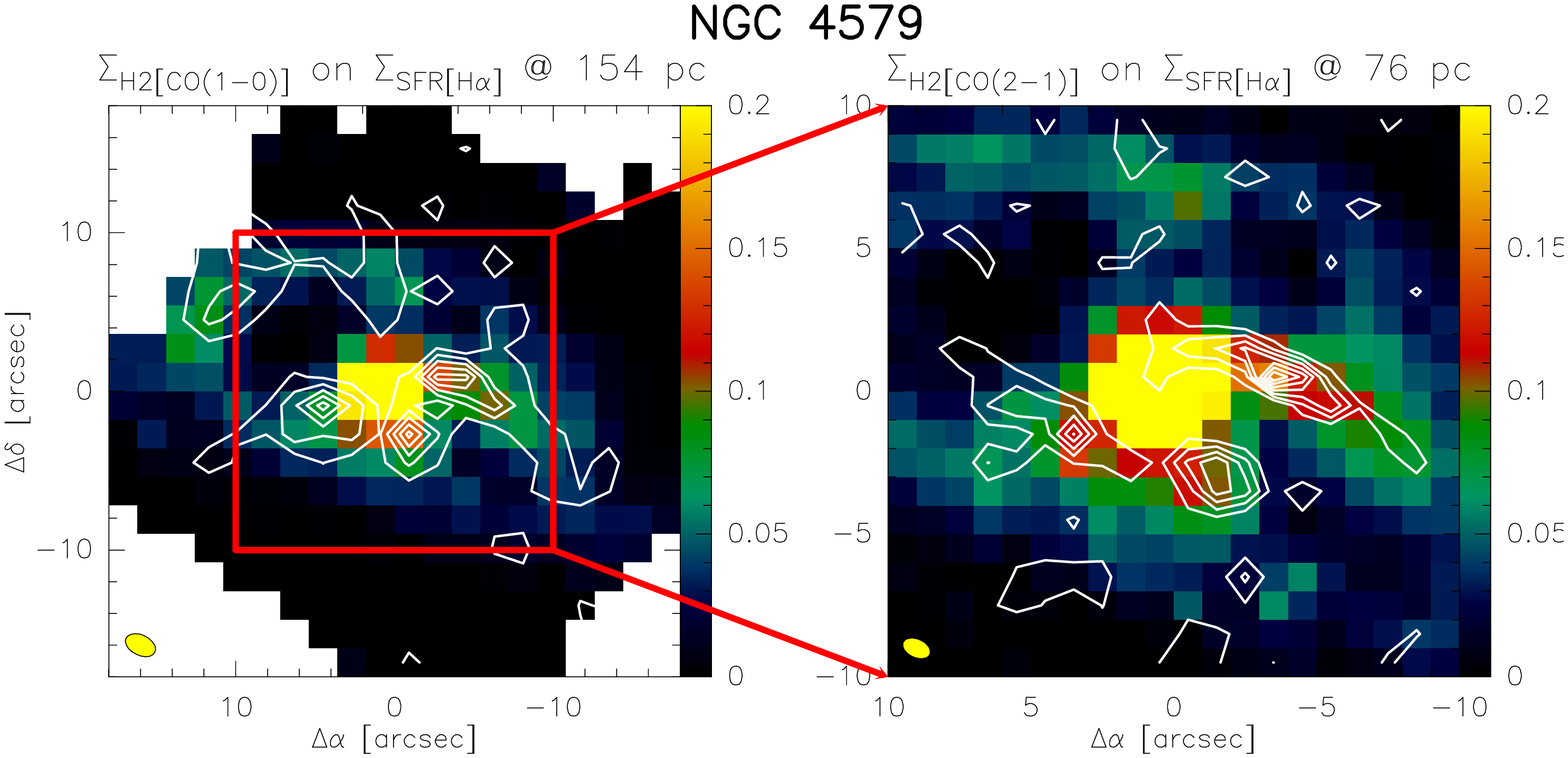}
\includegraphics[width=1\textwidth]{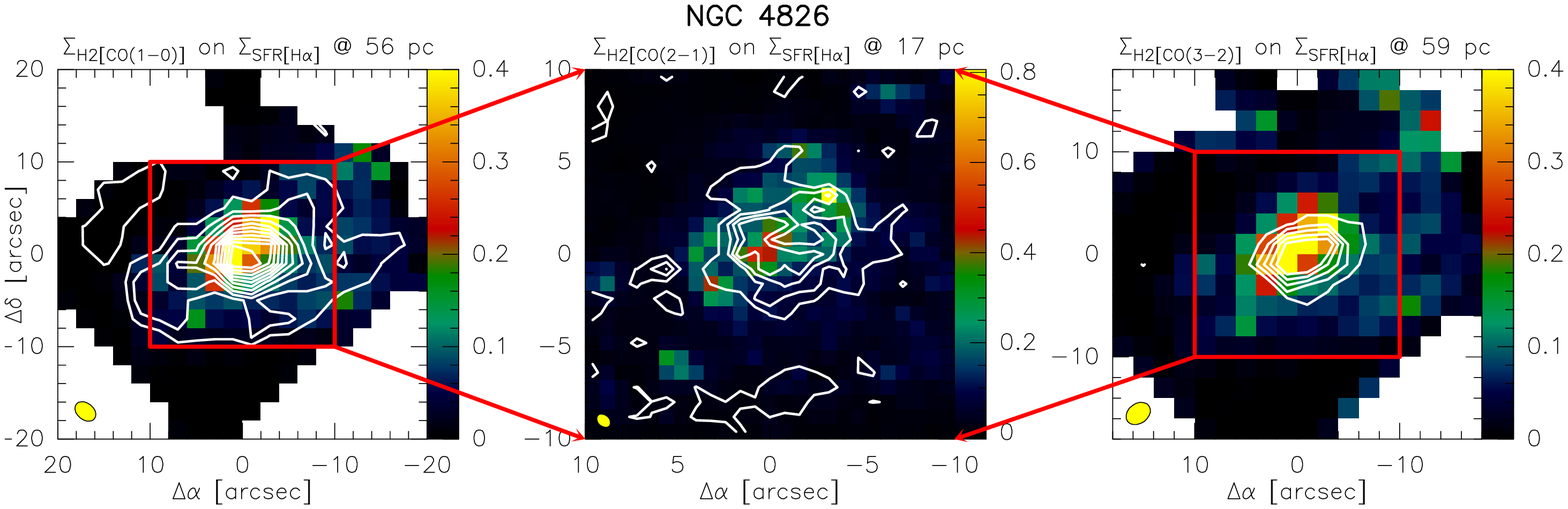}
\caption{
\sigmahtwo\ contours in M$_{\odot}$~pc$^{-2}$ derived from $^{12}$CO(1--0), $^{12}$CO(2--1), and
$^{12}$CO(3--2) overlaid on \sigmasfr\ images in M$_{\odot}$~yr$^{-1}$~kpc$^{-2}$ at the intrinsic 
spatial resolution of the $^{12}$CO data and with pixel sizes equal to the $^{12}$CO resolution. 
The beams are plotted (in yellow) in the lower left corners of maps and their values are listed in Table~\ref{tab:info-co}.
($\Delta \alpha $,  $\Delta \delta $)-offsets are with respect to the dynamical center of each galaxy.
\sigmahtwo\ contours are drawn starting from noise levels $>$3$\sigma$ with contour spacings 
that are whole multiples of $>$3$\sigma$.
}
\label{fig:overlays}
\end{figure*}

\section{The molecular star formation relation for individual galaxies}
\label{sec:sf-law}
In this section we present the results obtained by the study of the pixel-by-pixel molecular 
K-S relation for each galaxy of our sample, using the two or three lowest $^{12}$CO transitions for deriving 
\sigmahtwo\ and H$\alpha$ emission for \sigmasfr, with spatial resolution ranging from $\sim$20~pc to 200~pc. We show two sets of figures
for each galaxy.
The first one (Fig.~\ref{fig:overlays}) displays overlays of \sigmahtwo\ contours (above 3$\sigma$) in M$_{\odot}$~pc$^{-2}$ 
derived from $^{12}$CO(1--0), $^{12}$CO(2--1), and $^{12}$CO(3--2) emission line
on \sigmasfr\ maps in M$_{\odot}$~yr$^{-1}$~kpc$^{-2}$ derived from H$\alpha$ emission line at the resolution of original $^{12}$CO data.
The images involved in these overlays have been obtained following prescriptions
described in Sects.~\ref{sec:treatment}, \ref{sec:h2}, and \ref{sec:sfr}.
Each panel of Fig.~\ref{fig:overlays} reports the galaxy name, the reference to the $^{12}$CO emission line used to derive \sigmahtwo, and
the spatial resolution under analysis, i.e. the one offered by the intrinsic $^{12}$CO line map and at which the H$\alpha$ image 
has been degraded. 

\begin{figure*}
\begin{minipage}[b]{18cm}
\centering
\includegraphics[width=0.45\textwidth]{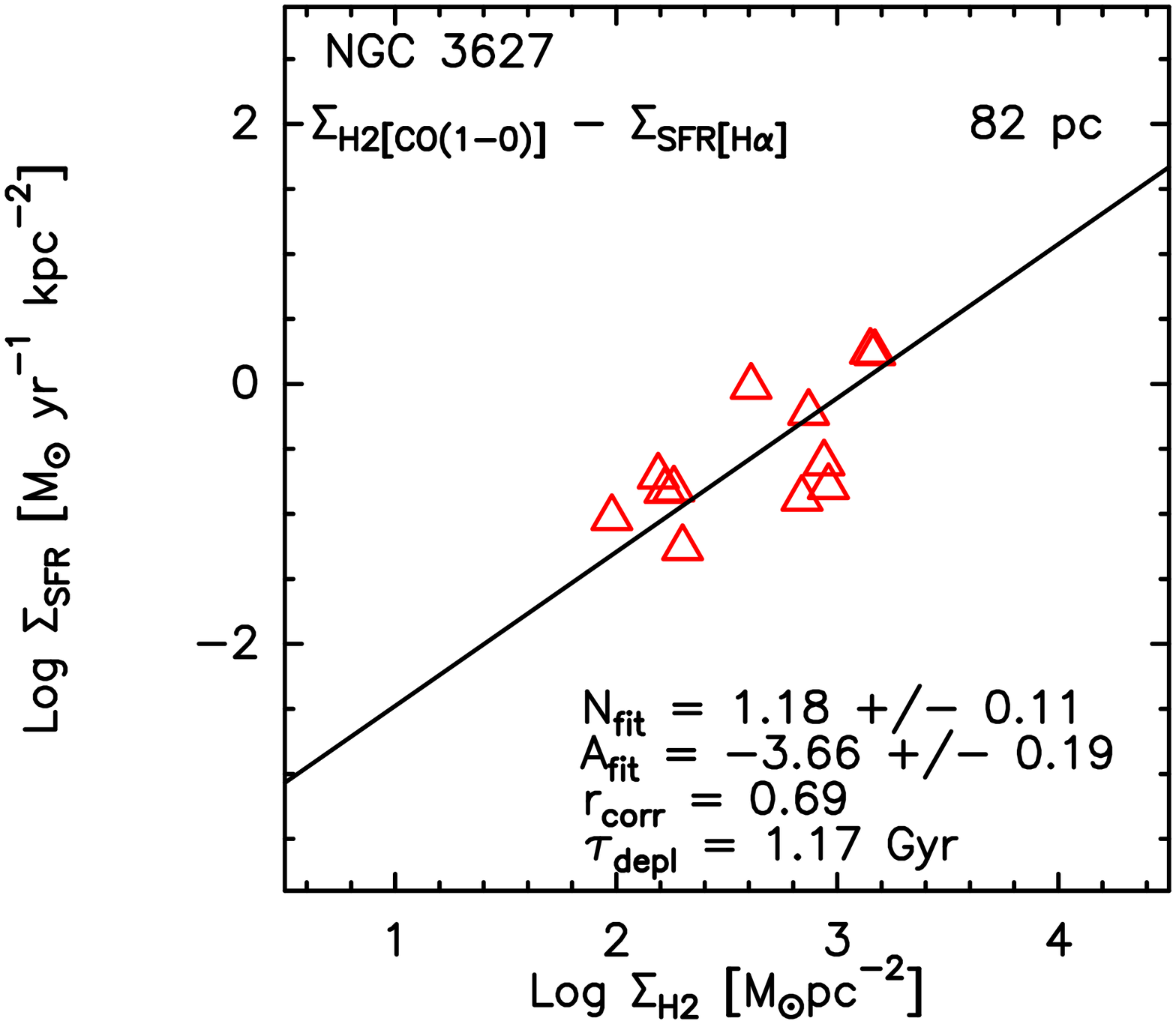}
\hspace{1cm}
\includegraphics[width=0.45\textwidth]{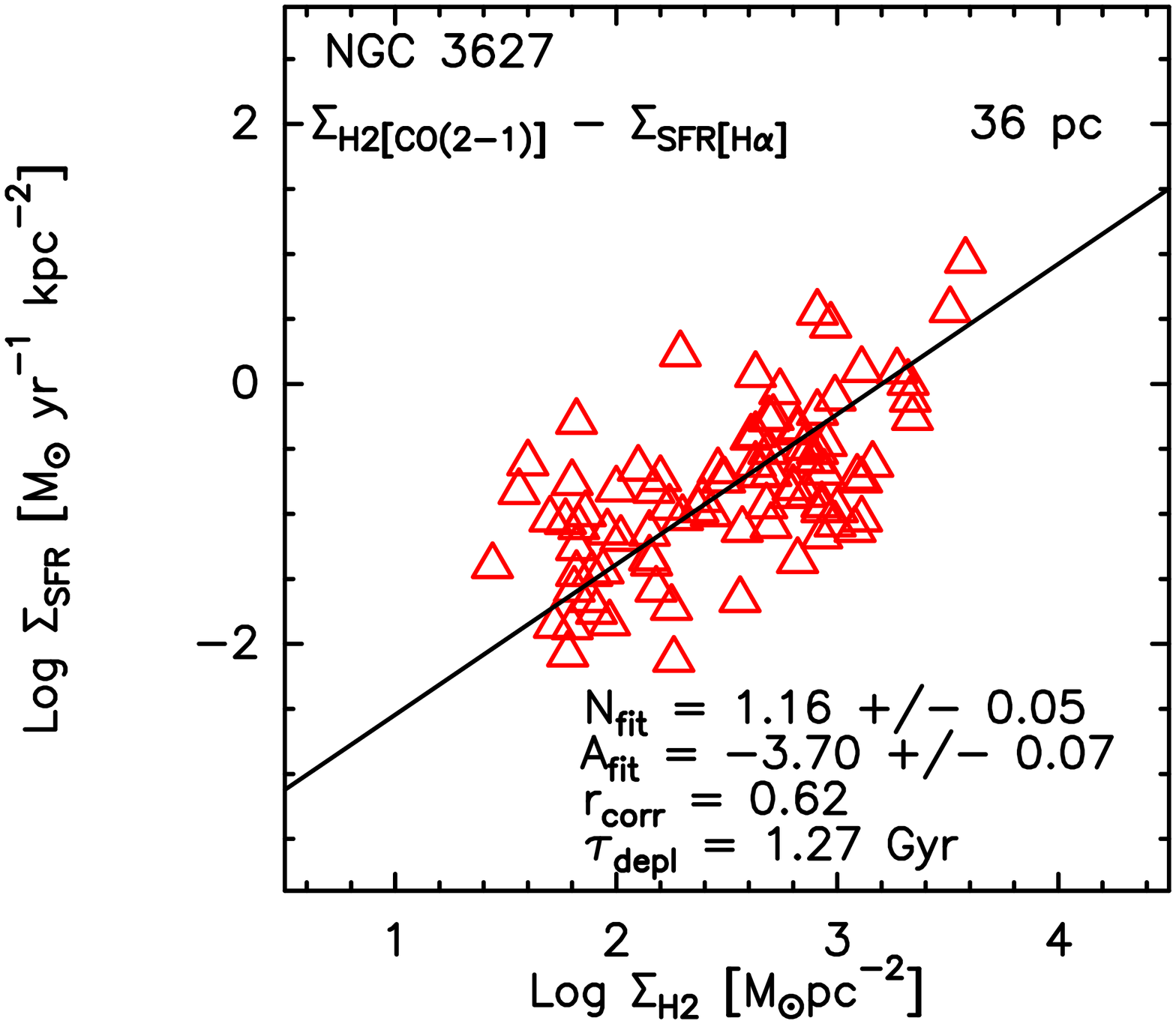}
\end{minipage}
\caption{\textit{Left panel}: 
The K-S relation plot for NGC~3627 at the resolution of 82~pc.
\sigmahtwo\ was derived from the $^{12}$CO(1--0) emission line map
and \sigmasfr\ from the H$\alpha$ image based on Eqs. (\ref{s10}) and 
(\ref{sfr-halpha}), respectively.
Red triangles indicate data points above 3$\sigma$ significance both in \sigmahtwo\ and \sigmasfr,
within a radius of 1.3~kpc (on the plane of the galaxy).
The solid black line indicates the OLS bisector fitting line (see Sect.~\ref{sec:fit} for fitting method).  
The index $N_{\rm fit}$ and the intercept $A_{\rm fit}$ of the OLS bisector fitting line, 
the Pearson correlation coefficient $r_{\rm core}$, and \taudepl\ values are reported in figure.  
\textit{Right panel}:
Same as left panel with \sigmahtwo\ derived from the $^{12}$CO(2--1) emission line map
based on Eq. (\ref{s21}) at the resolution of 36~pc.
Red triangles indicate data points above 3$\sigma$ significance both in \sigmahtwo\ and \sigmasfr,
within a radius of 1.1~kpc (on the plane of the galaxy).
\label{fig:ks-n3627}
}
\end{figure*}

\begin{figure*}
\begin{minipage}[b]{18cm}
\centering
\includegraphics[width=0.45\textwidth]{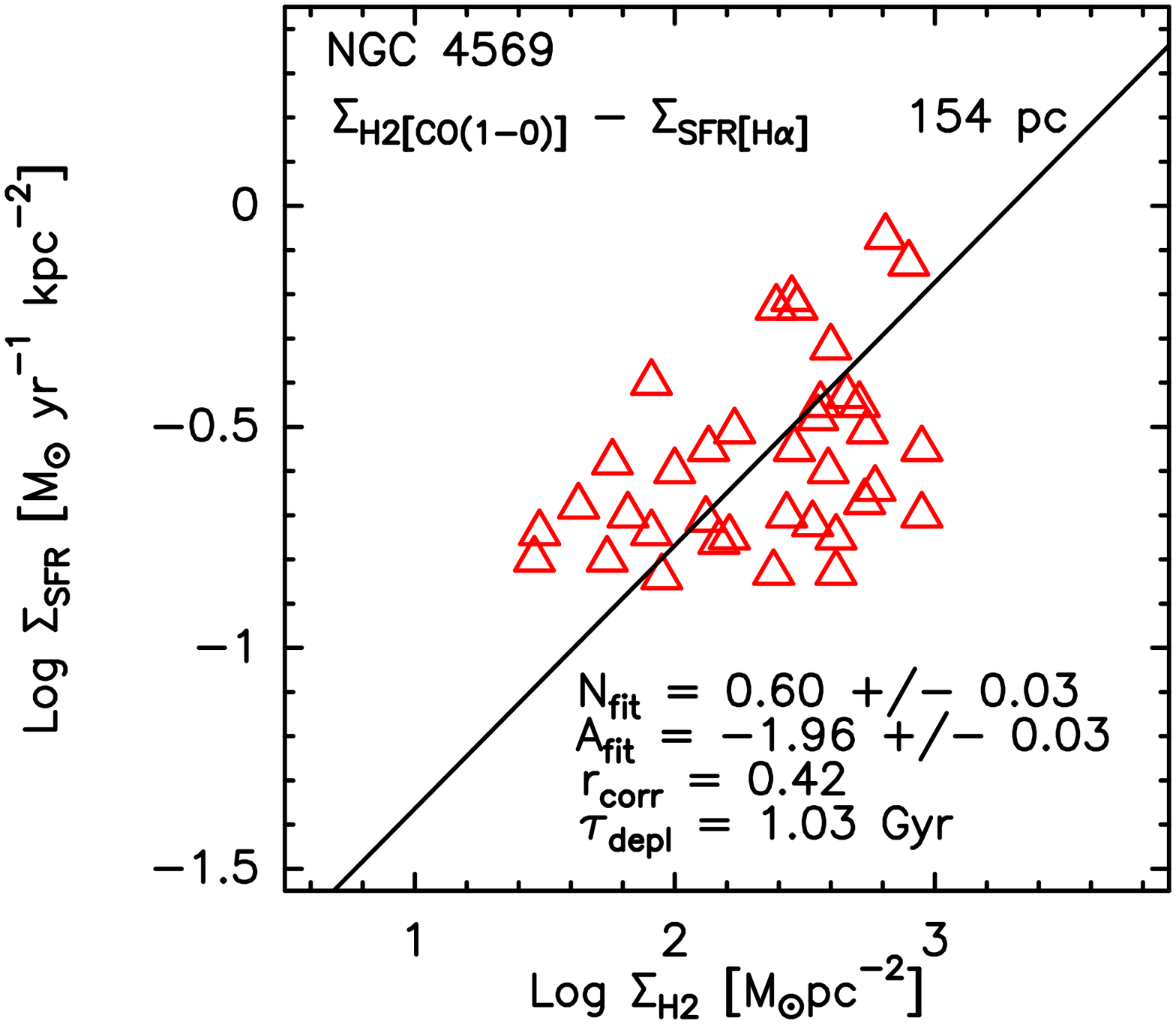}
\hspace{1cm}
\includegraphics[width=0.45\textwidth]{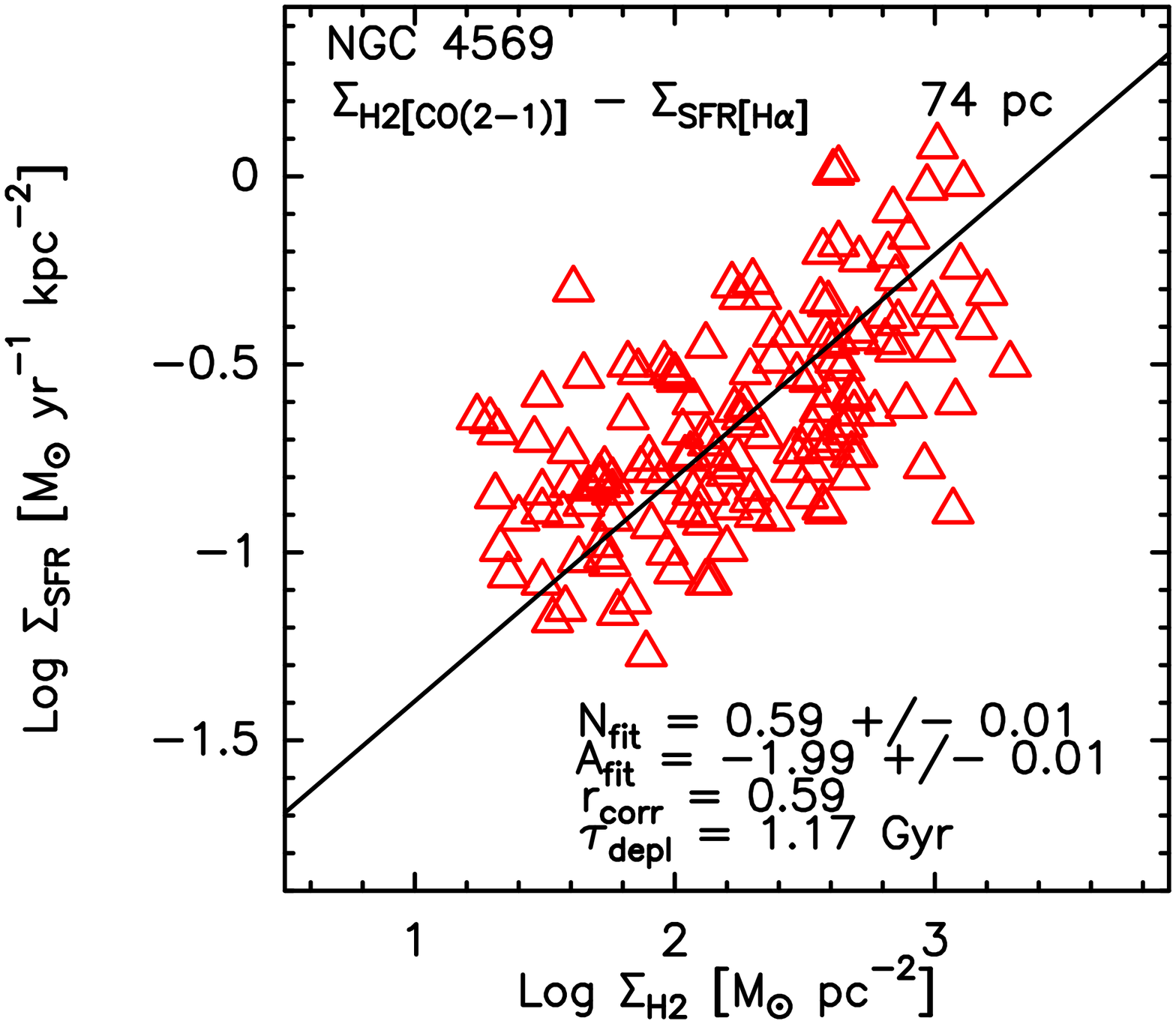}
\end{minipage}
\caption{\textit{Left panel}: 
Same as left panel of Fig.~\ref{fig:ks-n3627} for NGC~4569 at the resolution of 154~pc.
Red triangles indicate data points above 3$\sigma$ significance both in \sigmahtwo\ and \sigmasfr,
within a radius of 2.9~kpc (on the plane of the galaxy).
\textit{Right panel}: 
Same as left panel with \sigmahtwo\ derived from the $^{12}$CO(2--1) emission line map
at the resolution of 74~pc.
Red triangles indicate data points above 3$\sigma$ significance both in \sigmahtwo\ and \sigmasfr,
within a radius of 2.1~kpc (on the plane of the galaxy).
\label{fig:ks-n4569}
}
\end{figure*}

The second set of images (Figs.~\ref{fig:ks-n3627}--\ref{fig:co32}) consists of two or three plots for galaxy, 
according to the available $^{12}$CO line maps, displaying the K-S relation at the intrinsic resolution of 
the $^{12}$CO line data.
These plots contain the galaxy name, the reference to the adopted $^{12}$CO line for deriving \sigmahtwo, 
the spatial resolution, the fit line derived from the OLS bisector method described in Sect.~\ref{sec:fit}, 
the values of the power index $N_{\rm fit}$ and of the intercept $A_{\rm fit}$ and their uncertainties of the fit line according to 
equation (\ref{log}), and the derived value for $\tau_{\rm depl}$ defined as 
$\tau_{\rm depl} \equiv \langle \Sigma_{\rm H_2} \rangle / \langle \Sigma_{\rm SFR} \rangle$ 
\citep[e.g.,][]{leroy13}.

The available FoVs vary as a function of observed line and instrument.
IRAM $^{12}$CO(1--0) maps have a FoV of the primary beam of $\sim$42\arcsec, IRAM $^{12}$CO(2--1) of $\sim$21\arcsec,
SMA $^{12}$CO(3--2) of $\sim$36\arcsec, while $HST$ H$\alpha$ images have usable FoVs ranging from
$\sim$24\arcsec\ to $\sim$40\arcsec.

The results of this analysis, including the findings derived convolving all maps to the common spatial 
resolution of 200~pc, are collected in Table~\ref{tab:fits}. 
In this table, Col. (1) indicates the galaxy name, Col. (2) the $^{12}$CO transition
used to derive \sigmahtwo, Col. (3) the spatial resolution in pc, Col. (4) the available FoV (diameter)
in arcsec on the plane of the sky, Col. (5) the radius under investigation in kpc on the plane of the galaxy.
Columns (6) and (7) give the power index ($N_{\rm fit(OLS\,bis.)}$) and the intercept ($A_{\rm fit(OLS\,bis.)}$) 
and their uncertainties of the OLS bisector fitting line (for simplicity in the text these two parameters are indicated with 
$N_{\rm fit}$ and $A_{\rm fit}$). 
Column (8) indicates the Pearson correlation coefficient ($r_{\rm corr(OLS\,bis.)}$) of the OLS bisector fitting line 
(for simplicity in the text indicated with $r_{\rm corr}$) and the number of points under analysis (n. pts),
Cols. (9) and (10) the power index ($N_{\rm fit(RR)}$) and the intercept ($A_{\rm fit(RR)}$)  
and their uncertainties of the robust regression fitting line;,
Col. (11) the mean \sigmahtwo\ (<$\Sigma{\rm_{H2}}$>) in M$_{\odot}$\,pc$^{-2}$ within the FoV and taking only data points above 3$\sigma$ significance 
into account, Col. (12) the mean \sigmasfr\ (<$\Sigma{\rm_{SFR}}$>) 
in M$_{\odot}$\,yr$^{-1}$\,kpc$^{-2}$ under the same conditions, Col. (13) the molecular \taudepl\ 
in Gyr, and Col. (14) the final pixel size of $^{12}$CO and H$\alpha$ images after procedures described 
in Sect.~\ref{sec:treatment}.

\subsection{NGC~3627}
\label{sec:n3627}
NGC~3627 (Messier 66) is an interacting \citep[e.g.,][]{vivi04} and barred galaxy
classified as SAB(s)b at a distance of 10~Mpc, with signatures of a LINER/Seyfert 2 type nuclear activity \citep{ho97}.
With NGC~3623 and NGC~3628, it forms the well-known Leo Triplet (M~66 Group, VV~308). 
Optical broad-band images of NGC~3627 reveal a pronounced and asymmetric spiral pattern
with heavy dust lanes, indicating strong density wave action \citep{ptak06}. 
While the western arm is accompanied by weak traces of SF visible in H$\alpha$, the eastern arm
contains a star-forming segment in its inner part \citep{smith94,chemin03}.
NGC~3627 also possesses X-ray properties of a galaxy with a recent starburst \citep{dahlem96}. 
Both the radio continuum \citep[2.8~cm and 20~cm,][]{urbanik85,paladino08} and the CO 
emissions \citep[e.g.,][]{regan01,paladino08,haan09,vivi11,watanabe11} show a nuclear peak, 
extending along the leading edges of the bar that forms two broad maxima at the bar ends, and then the spiral arms trail
off from the bar ends. 
In contrast, the \hi\ emission exhibits a spiral morphology without signatures of a bar in the atomic gas
\citep{haan08,walter08}.
The derived gravity torque budget shows that NGC~3627 is a potential \textit{\emph{smoking gun}} of inner gas inflow at a 
resolution of $\sim$60~pc \citep[][]{vivi11}.
In addition to $^{12}$CO lines,  other molecular transitions have been detected in
NGC~3627, including HCN(1--0), HCN(2--1), HCN(3--2), HCO$^{+}$(1--0), and HCO$^{+}$(3--2), suggesting 
the presence of high density gas \citep[e.g.,][]{gao04,krips08}.

The panels of the first line of Fig.~\ref{fig:overlays} show the superposition of \sigmahtwo\ image contours derived 
from $^{12}$CO(1--0) and $^{12}$CO(2--1) emission lines overlaid on $\Sigma{\rm_{SFR}}$ images estimated from the 
H$\alpha$ emission at the intrinsic resolutions of CO maps, i.e., $\sim$1\farcs7 ($\sim$82~pc) for the (1--0) transition 
and $\sim$0\farcs7 ($\sim$36~pc) for the (2--1) one. The value of
\sigmahtwo\ has been derived from $^{12}$CO(1--0) and $^{12}$CO(2--1) emission lines
by using Eqs.~(\ref{s10}) and (\ref{s21}), respectively, and \sigmasfr\ by using Eq.~(\ref{sfr-halpha}).
For $^{12}$CO(1--0) (left panel), the \sigmahtwo\ and \sigmasfr\ peaks are spatially coincident in the limit of the 
resolution of $\sim$1\farcs7 and the two distributions are quite consistent within a radius of 
$\sim$3--4\arcsec\ (150--200\,pc) from the nucleus (on the plane of the sky). 
At larger distances, \sigmahtwo\ and \sigmasfr\ are not correlated, 
mainly because \sigmasfr\ is not distributed along a bar as is \sigmahtwo. 
Similar (anti-)correlations also characterize the comparison between \sigmahtwo\ derived from $^{12}$CO(2--1) 
emission line and \sigmasfr\ 
at the resolution of $\sim$0\farcs7 (right panel).

The lefthand panel of Fig.~\ref{fig:ks-n3627} shows the molecular K-S relation derived for NGC~3627 using 
the $^{12}$CO(1--0) emission line to estimate \sigmahtwo\ at the resolution of 82~pc.
At this resolution the power index 
$N_{\rm fit}$ is equal to $1.18 \pm 0.11$ 
within a radius of 1.3~kpc (on the plane of the galaxy).
Following the same procedure but convolving the original $^{12}$CO(1--0) and H$\alpha$ maps at the 
lower resolution of 200~pc, the K-S relation has  
$N_{\rm fit}$ equal to $1.11 \pm 0.41$.
The righthand panel of Fig.~\ref{fig:ks-n3627} shows the results obtained using the $^{12}$CO(2--1) emission line to derive 
\sigmahtwo (Eq. {\ref{s21}}) at the resolution of 36~pc.
At this resolution $N_{\rm fit}  = 1.16 \pm 0.05$
within a radius of 1.1~kpc (on the plane of the galaxy), while 
$N_{\rm fit}$ increases to $1.59 \pm 0.63$
at the resolution of 200~pc.
Although all the derived $N_{\rm fit}$ values are consistent with literature results (see later discussion in Sect.~\ref{sec:sf-index}), 
the K-S relations studied at the resolution of 200~pc -- for both $^{12}$CO(1--0) and $^{12}$CO(2--1) 
emission lines -- with only four data points involved in the analysis do not allow us to infer statistical conclusions.
In NGC~3627, we can only say that $N_{\rm fit} \approx 1.2$ both with \sigmahtwo\ derived from $^{12}$CO(1--0) 
at 82~pc and from $^{12}$CO(2--1) at 36~pc.

Neglecting the 200~pc-resolution cases, the Pearson correlation coefficient is $\sim$0.6 -- 0.7
and \taudepl\ is $\sim$1.2 -- 1.3~Gyr for both the lowest $^{12}$CO transitions.

\subsection{NGC~4569}
\label{n4569}
NGC~4569 (Messier~90) is a bright SAB(rs)ab galaxy at a distance of 17~Mpc in the Virgo cluster. 
A large scale bar is seen in NIR images \citep[][]{laurikainen02} and is almost aligned with the major axis of
the galaxy \citep[PA = 15 deg according to][]{jogee05}. 
The galaxy harbors a nucleus of the transition type \citep[type T2 in][]{ho97},
which exhibits a pronounced nuclear starburst activity.
\hi\ emission line observations have revealed that in NGC~4569 the atomic gas is distributed along 
a central bar with radius of $\sim$60\arcsec\ \citep[$\sim$5~kpc,][]{haan08}.
Interferometric CO observations of NGC~4569 presented previously 
\citep[e.g.,][]{helfer03,jogee05,nakanishi05,boone07} 
have shown that the major part of the molecular gas detected in the inner 20$^{\prime\prime}$ is concentrated 
within a radius of 800~pc, distributed along the large scale stellar bar seen in NIR observations,
and with a peak close to the center and another one at $\sim$500~pc from it.
A hole in the CO distribution coincides with the nucleus where most of the H$\alpha $ emission and blue light are emitted
\citep[][]{pogge00}.
\citet{boone07} also demonstrate that the gravitational torques are able to efficiently funnel the gas down to $\sim$300~pc.

The panels of the second line of Fig.~\ref{fig:overlays} show the comparison between  
\sigmahtwo\ and \sigmasfr\ distributions in NGC~4569 at resolutions of $\sim$1\farcs9 ($\sim$154~pc from $^{12}$CO(1--0)),
$\sim$0\farcs9 ($\sim$74~pc from $^{12}$CO(2--1)), and $\sim$2\farcs3 ($\sim$190~pc from $^{12}$CO(3--2)).
The case involving the $^{12}$CO(3--2) emission line refers to Eq. ({\ref{s32}) for deriving 
\sigmahtwo.
>From these overlays it can be seen that \sigmahtwo\ and \sigmasfr\ are differently distributed at the available resolutions. 
While \sigmasfr\ is centrally concentrated and peaked, \sigmahtwo\ (derived both from $^{12}$CO(1--0) and
$^{12}$CO(2--1) emission line) is distributed along a large scale bar of $\sim$17\arcsec\ $\times$ 6\arcsec\ 
($\sim$1.4~kpc~$\times$~0.5~kpc) in size (in the plane of the sky) whose peak is far away from the peak of \sigmasfr. 
The same distribution is also visible in the SMA $^{12}$CO(3--2) map.
There are, however, differences, perhaps due to the lower S/N ratio of the $^{12}$CO(3--2) SMA data 
relative to the IRAM observations \citep[][]{boone11}.

Similar to NGC~3627, NGC~4569 exhibits a centrally concentrated morphology in the 
\sigmasfr\ images and a bar-like distribution in the $\Sigma_{\rm gas}$ maps.
This could mean that the action of the bar has transported the gas to the
nuclear regions to fuel a mini-starburst episode as found in many nearby spiral galaxies \citep{sakamoto99}.

Figure \ref{fig:ks-n4569} and the lefthand panel of Fig.~\ref{fig:co32} 
show the molecular K-S relation derived for NGC~4569 with \sigmahtwo\ estimated from the three 
lowest $^{12}$CO emission lines at the intrinsic resolution of the $^{12}$CO maps.
These figures, together with findings collected in Table~\ref{tab:fits}, show that 
NGC~4569 has an index $N_{\rm fit}$ of the K-S relation that is sublinear ($\sim$0.6 -- 0.7) 
for the three available $^{12}$CO transitions studied at spatial scales from 74 to 200~pc 
(producing six subcases), roughly constant as a function of resolution for a given $^{12}$CO transition, and
only slightly varying as a function of $^{12}$CO transition at a resolution of 200~pc.

For NGC~4569, the Pearson correlation coefficient is approximately invariant with respect 
to the resolution, from 74 to 200~pc, for a given $^{12}$CO transition.
In contrast to this, $r_{\rm corr}$ varies as a function of $^{12}$CO line at a given resolution 
(i.e., 200~pc).
The best $r_{\rm corr}$ is obtained with the $^{12}$CO(2--1) line ($\sim$0.5--0.6),
while the worst one with the $^{12}$CO(3--2) transition ($\sim$0.3) ($^{12}$CO(1--0) gives 
$r_{\rm corr}$~$\sim$0.4).

As in NGC~3627, NGC~4569 has a short molecular \taudepl\ of $\sim$1~Gyr, suggesting 
that the gas is efficiently converted in stars.
Within a radius of 0.5~kpc, \taudepl\ is even smaller assuming values of $\sim$0.7--0.9~Gyr.

\begin{figure*}
\begin{minipage}[b]{18cm}
\centering
\includegraphics[width=0.45\textwidth]{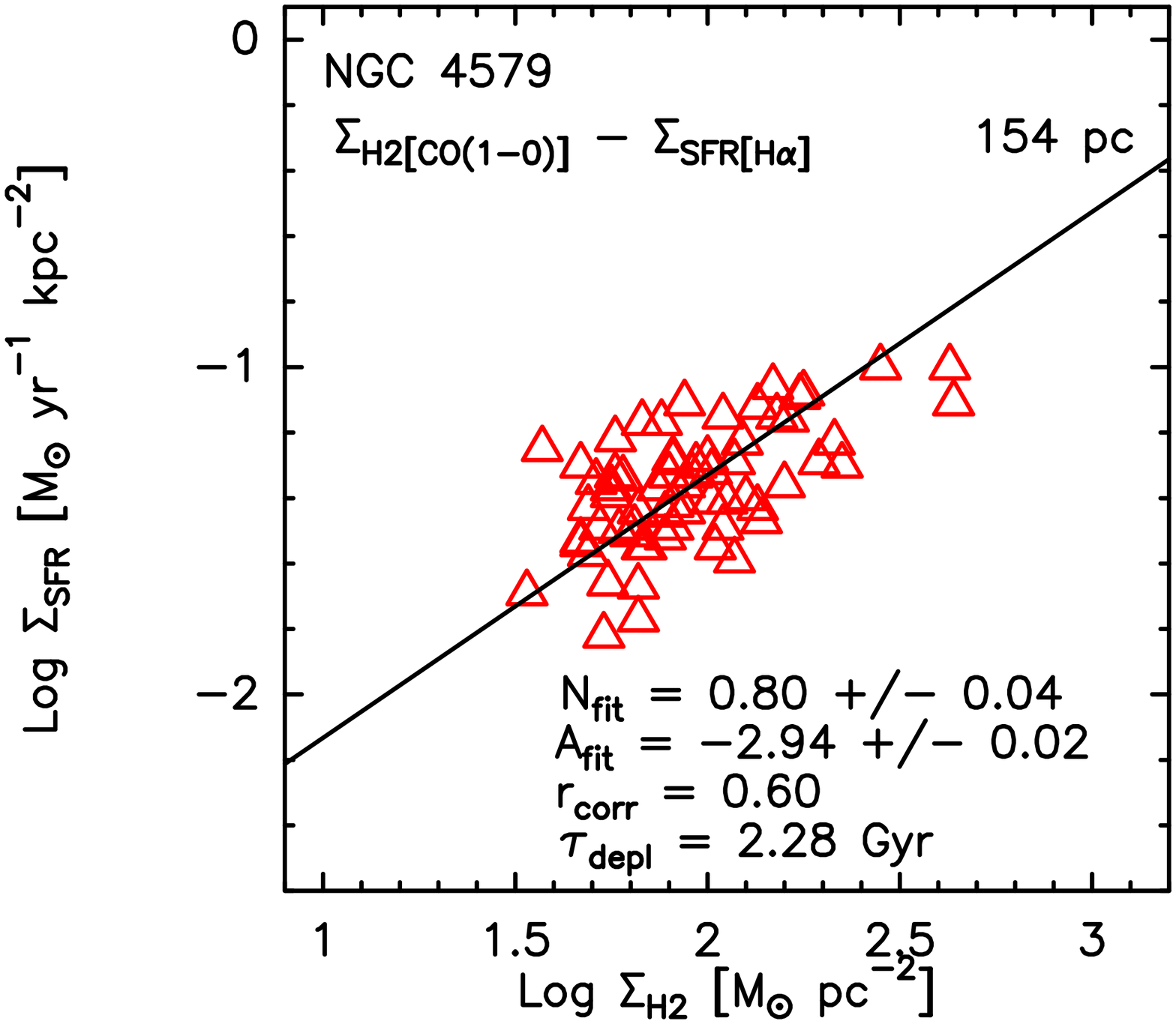}
\hspace{1cm}
\includegraphics[width=0.45\textwidth]{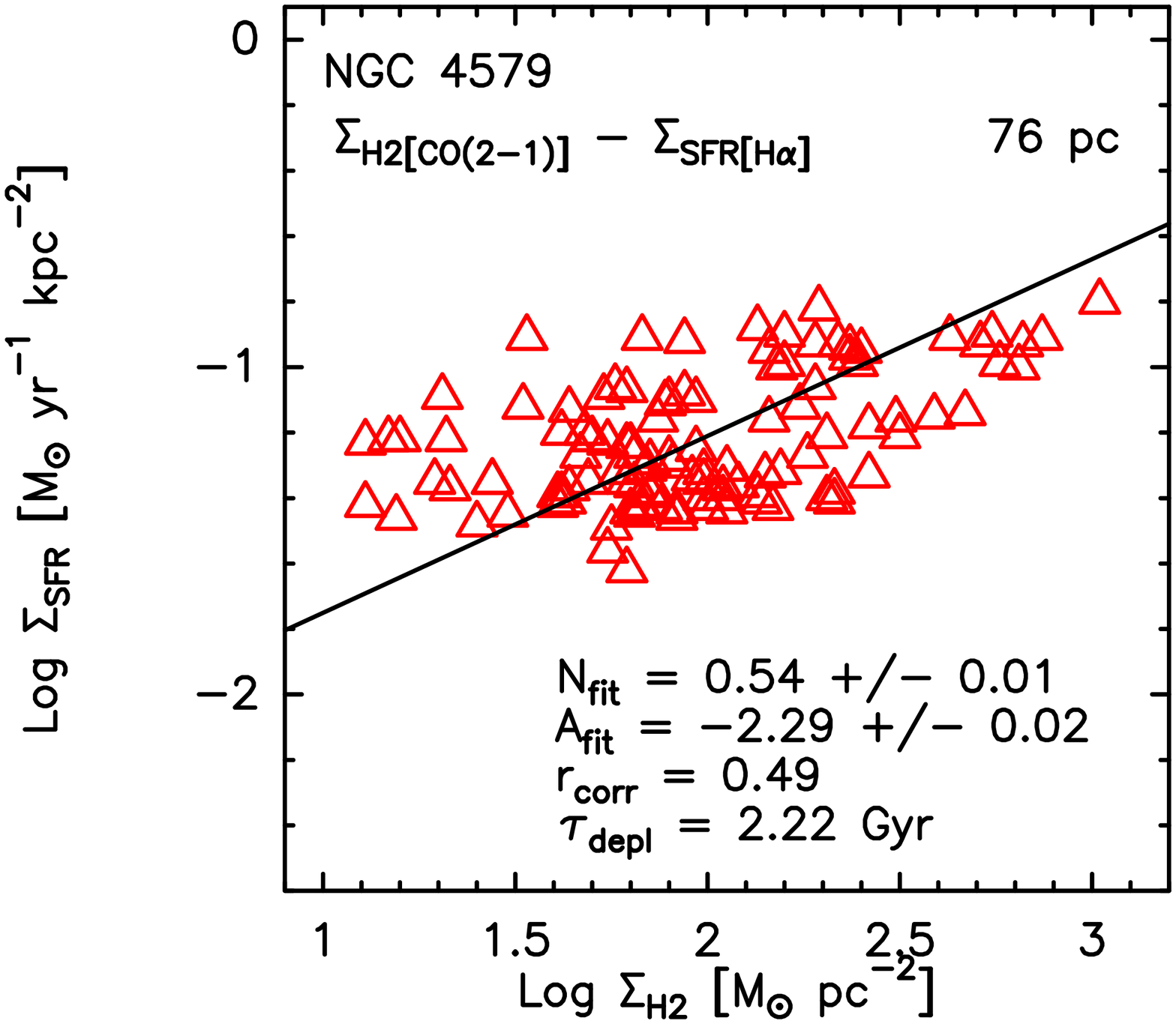}
\end{minipage}
\caption{\textit{Left panel}: 
Same as left panel of Fig.~\ref{fig:ks-n3627} for NGC~4579 at the resolution of 154~pc.
Red triangles indicate data points above 3$\sigma$ significance both in \sigmahtwo\ and \sigmasfr,
within a radius of 2.7~kpc (on the plane of the galaxy).
\textit{Right panel}: 
Same as left panel with \sigmahtwo\ derived from the $^{12}$CO(2--1) emission line map
at the resolution of 76~pc.
Red triangles indicate data points above 3$\sigma$ significance both in \sigmahtwo\ and \sigmasfr,
within a radius of 1.5~kpc (on the plane of the galaxy).
\label{fig:ks-n4579}
}
\end{figure*}

\subsection{NGC 4579}
\label{sec:n4579}
NGC~4579 (Messier~58) is a SAB(rs)b galaxy classified as an intermediate type 1 object (LINER/Seyfert 1.9) by \citet{ho97} at 
a distance of 20~Mpc in the Virgo cluster.
It also has an unresolved nuclear hard X-ray (variable) source with a prominent broad Fe K${\alpha }$ line 
\citep[][]{terashima00,ho01,eracleous02,dewangan04}.
A non-thermal radio continuum source is detected at the position of the AGN 
\citep[][]{hummel87,ho01b,ulvestad01,krips07}.
The NIR K-band image of NGC~4579 has revealed a large-scale stellar bar and a weak nuclear oval \citep{santi09}.
The 21~cm \hi\ line observations have shown that the atomic gas in NGC~4579 is currently piling up in a pseudo-ring 
(radius of $\sim$40\arcsec, $\sim$4~kpc) formed by two winding spiral arms that are morphologically decoupled 
from the bar structure \citep{haan08}.   
Molecular gas in the inner r $\leq $ 2 kpc disk is distributed in two spiral arms, an outer arc and a central lopsided disk-like structure
\citep[][]{santi09}. 
The derived gravity torque budget in NGC~4579 have shown that inward gas flow is occurring on different spatial scales in the disk,
with clear \textit{\emph{smoking gun}} evidence of inward gas transport down to r $\sim50$~pc.

The panels of the third line of Fig.~\ref{fig:overlays} show that} in NGC~4579,
\sigmahtwo\ and \sigmasfr\ have completely different distributions 
distributions at resolutions of both
$\sim$1\farcs6 ($\sim$154~pc from $^{12}$CO(1--0)) and $\sim$0\farcs8 ($\sim$76~pc from $^{12}$CO(2--1)).
The distribution of
\sigmasfr\ is centrally concentrated within the inner $\sim$4\arcsec\ (on the plane of the sky) 
and, at larger distances from the center, lies along two spiral arms extending up to 
$\sim$12\arcsec\ from the nucleus. 
The morphology of \sigmahtwo\ traced by $^{12}$CO(1--0) and $^{12}$CO(2--1) 
is instead mainly defined by two highly contrasted spiral lanes without a central peak \citep[][]{santi09}. 

Figure \ref{fig:ks-n4579} shows the results obtained from the analysis of the K-S relation for NGC~4579. The value of
$N_{\rm fit}$ ranges from $\sim$0.5 to $\sim$1.1 on spatial scales of 76--200~pc and 

considering $^{12}$CO(1--0) and $^{12}$CO(2--1) lines for \sigmahtwo\ derivation.
For a given $^{12}$CO line, $N_{\rm fit}$ decreases with finer resolution.
While for $^{12}$CO(1--0) $N_{\rm fit}$ gradually decreases from $\sim$0.9 at 200~pc-resolution to 
$\sim$0.8 at 154~pc-resolution, for $^{12}$CO(2--1) the decreasing of $N_{\rm fit}$ is stronger, from 1.1 to 0.5, 
possibly because the change from 200 to 76~pc in resolution is more drastic. 
At the common resolution of 200~pc, $N_{\rm fit}$ increases with higher $J$-CO transition.

In NGC~4579, the quality of the correlation tends to worsen with resolution.
While for $^{12}$CO(1--0) $r_{\rm corr}$ is approximately constant ($\sim$0.6) through 154--200~pc resolution, 
for $^{12}$CO(2--1) it drops down from 0.6 to 0.5 in the resolution range from 200 to 76~pc.
The correlation coefficient $r_{\rm corr}$ 
does not seem to depend on the CO transition,
since it is $\sim$0.6 
for both $^{12}$CO lines at the common resolution of 200~pc.

Unlike NGC~3627 and NGC~4569, NGC~4579 has a more ``standard'' molecular \taudepl\ of $\sim$2~Gyr
(see later discussion in Sect.~\ref{sec:depletion}), without a radial trend.
The similar trends obtained for $N_{\rm fit}$ and $r_{\rm corr}$ as a function of the resolution and for both 
the two lowest $^{12}$CO transitions suggest that in NGC~4579 the K-S relation is almost invariant 
with respect to $^{12}$CO(1--0) and $^{12}$CO(2--1) emission lines, but it changes  
slightly with spatial resolution.

\begin{figure*}
\begin{minipage}[b]{18cm}
\centering
\includegraphics[width=0.45\textwidth]{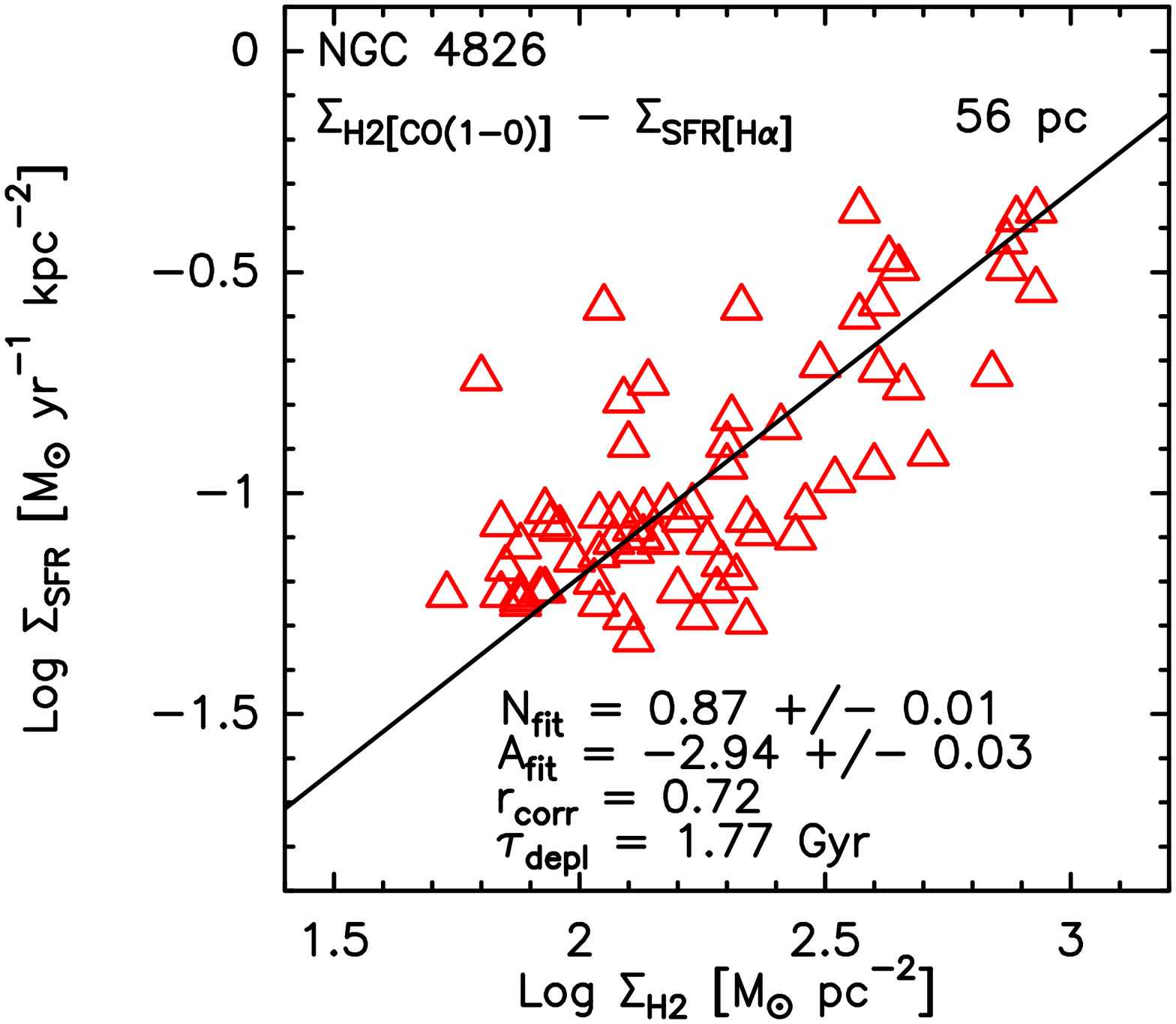}
\hspace{1cm}
\includegraphics[width=0.45\textwidth]{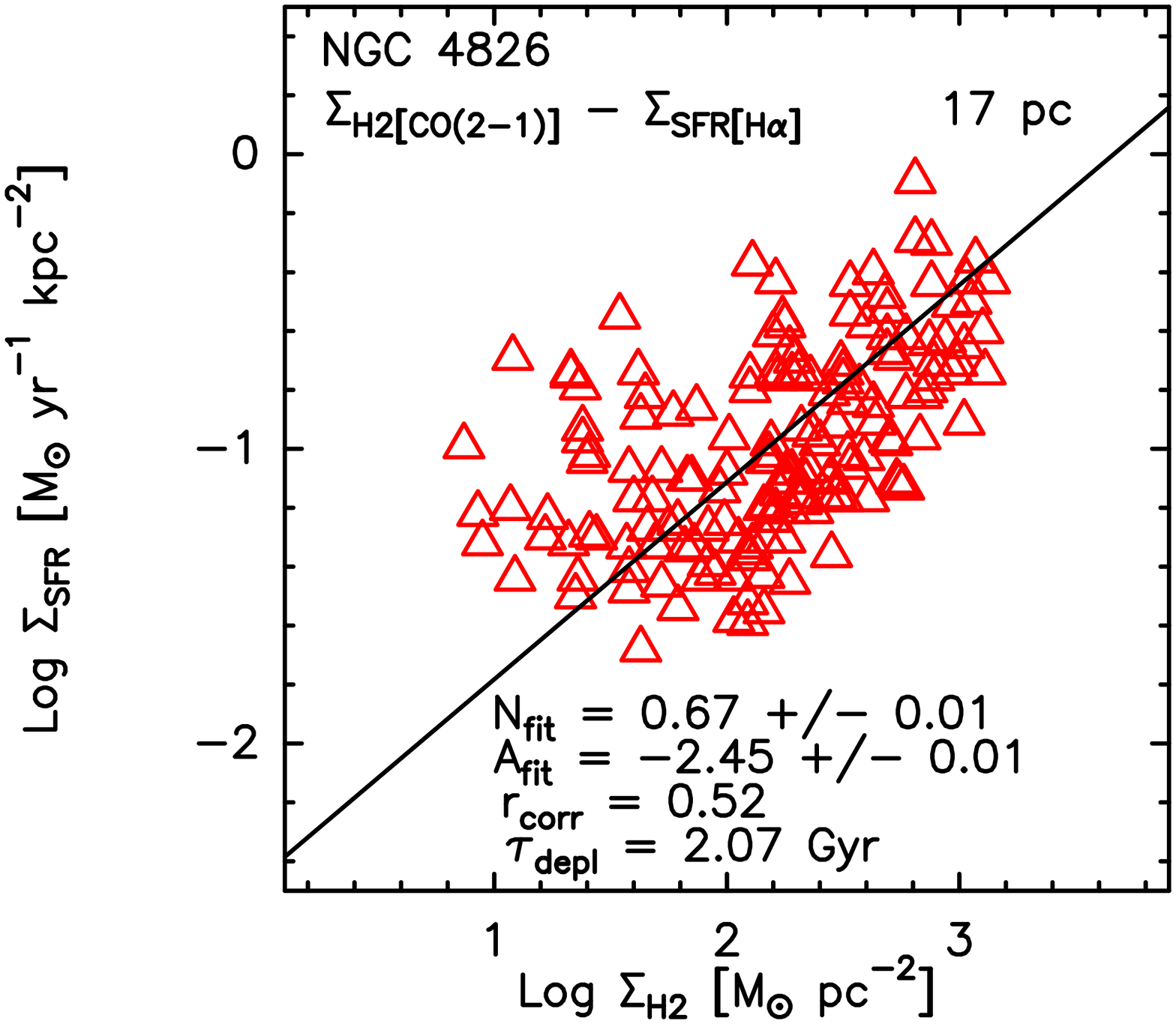}
\end{minipage}
\caption{\textit{Left panel}: 
Same as left panel of Fig.~\ref{fig:ks-n3627} for NGC~4826 at the resolution of 56~pc.
Red triangles indicate data points above 3$\sigma$ significance both in \sigmahtwo\ and \sigmasfr,
within a radius of 1.1~kpc (on the plane of the galaxy).
\textit{Right panel}: 
Same as left panel with \sigmahtwo\ derived from the $^{12}$CO(2--1) emission line map
at the resolution of 17~pc.
Red triangles indicate data points above 3$\sigma$ significance both in \sigmahtwo\ and \sigmasfr,
within a radius of 0.6~kpc (on the plane of the galaxy).
\label{fig:ks-n4826}
}
\end{figure*}

\begin{figure*}
\begin{minipage}[b]{18cm}
\centering
\includegraphics[width=0.45\textwidth]{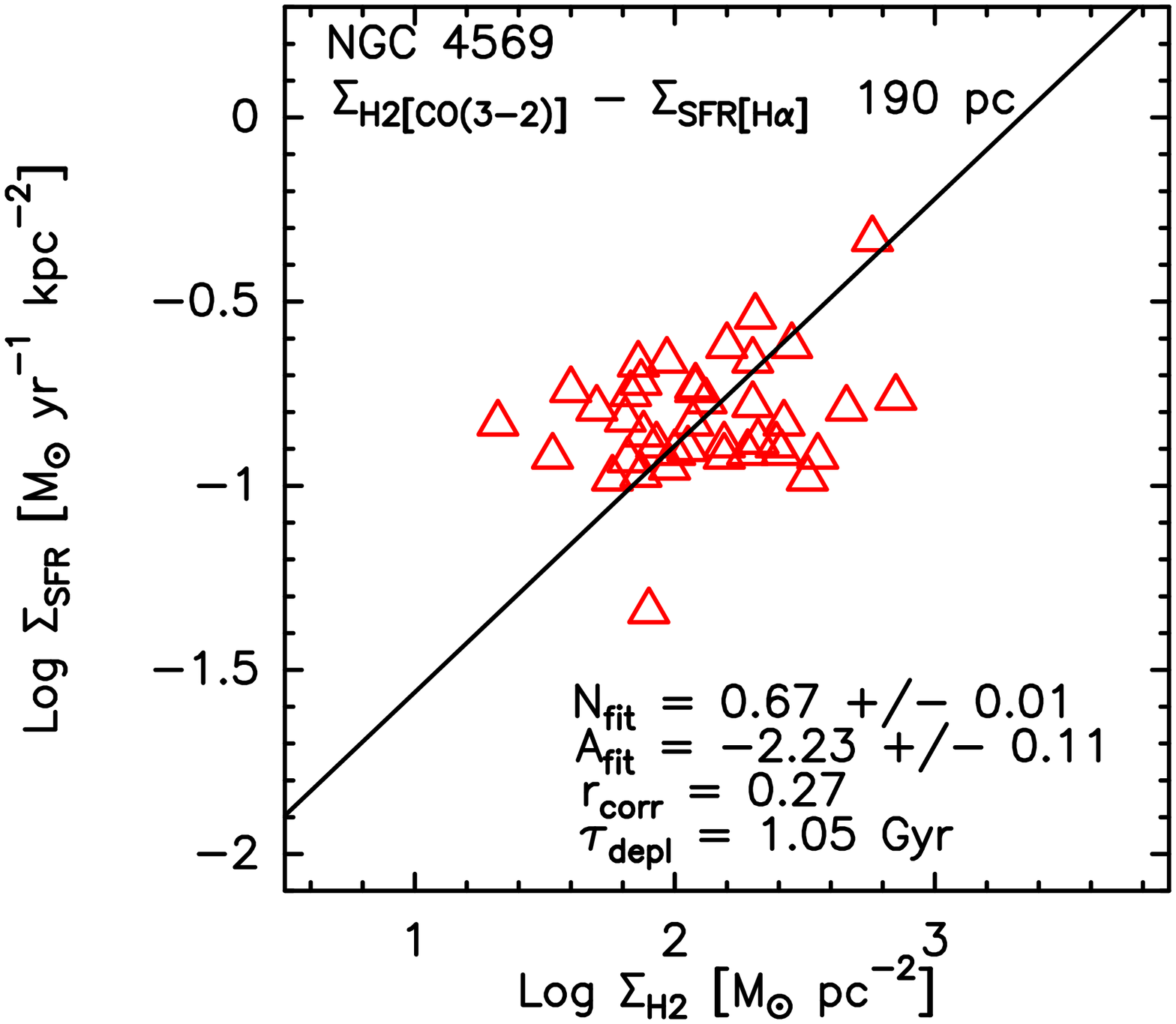}
\hspace{1cm}
\includegraphics[width=0.45\textwidth]{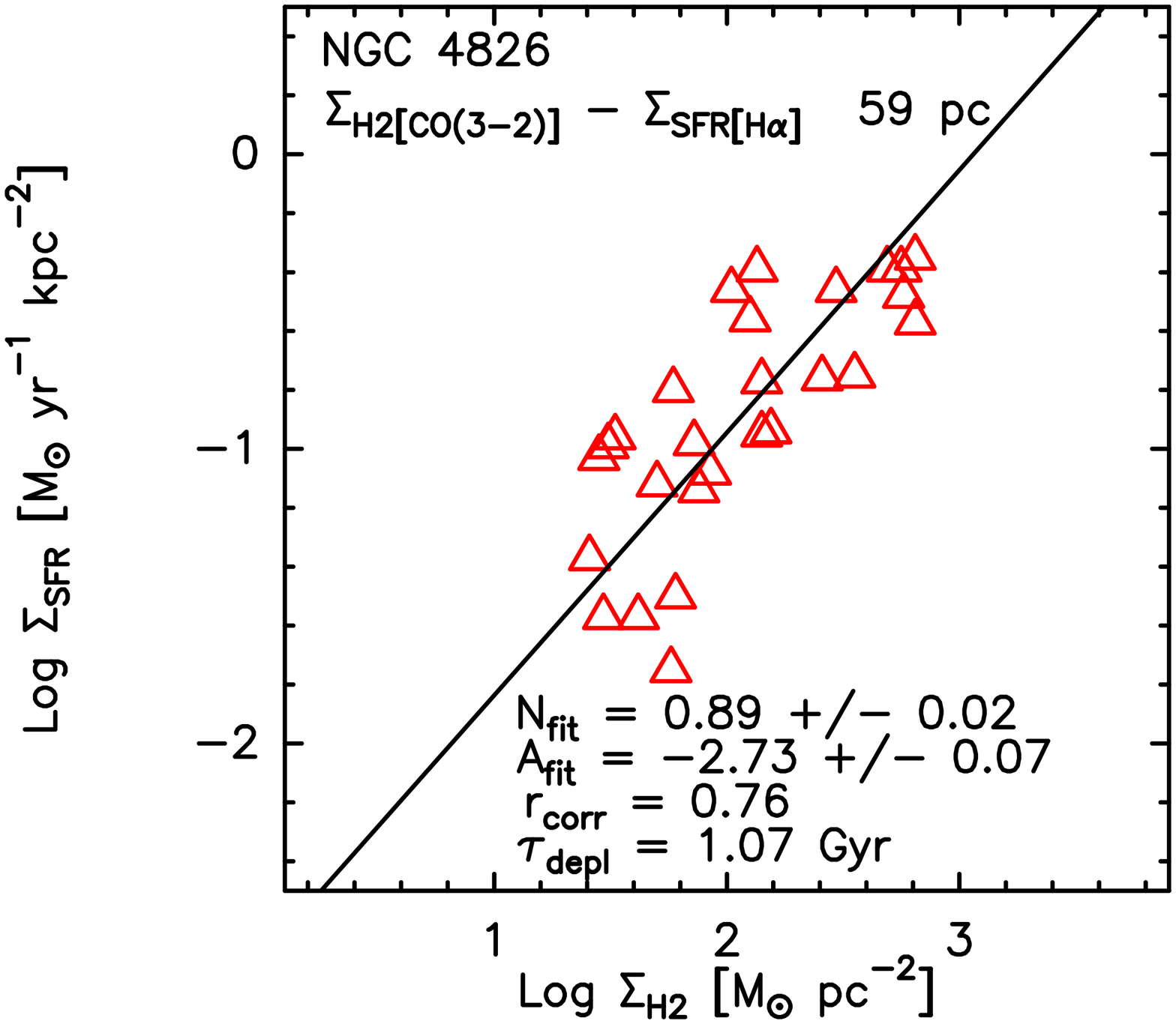}
\end{minipage}
\caption{\textit{Left panel}: 
Same as Fig.~\ref{fig:ks-n4569}, for NGC~4569, with \sigmahtwo\ derived from the $^{12}$CO(3--2) emission line map
based on equation (\ref{s32}) at the resolution of 190~pc.
Red triangles indicate data points above 3$\sigma$ significance both in \sigmahtwo\ and \sigmasfr,
within a radius of 2.9~kpc (on the plane of the galaxy).
\textit{Right panel}: 
Same as left panel for NGC~4826 at the resolution of 59~pc.
Red triangles indicate data points above 3$\sigma$ significance both in \sigmahtwo\ and \sigmasfr,
within a radius of 1.0~kpc (on the plane of the galaxy).
\label{fig:co32}
}
\end{figure*}

\subsection{NGC 4826}
\label{sec:n4826}
NGC~4826, also known as the ``Black Eye'' or ``Evil Eye'' galaxy due to its optical appearance, is the closest target ($\sim$5~Mpc) 
of the NUGA core sample, and its nucleus is classified as a LINER type \citep[][]{ho97}.
It hosts two nested counter-rotating atomic and molecular gas disks of comparable mass \citep[$\sim$10$^{8}$~M$_{\odot}$,][]{braun92,casoli93,braun94}.
The inner disk has a radius of $\sim$50\arcsec ($\sim$1.3~kpc), while the outer one from  $\sim$80\arcsec\ to  $\sim$9\farcm8
($\sim$2.1 -- 15.3~kpc). By studying the stellar kinematics along the principal axes of NGC~4826,
\citet{rix95} found that the stars rotate 
at all radii in the same direction as the inner disk, providing strong evidence that stars and gas are coplanar. 
NUGA observations have shown a high concentration of molecular gas (in $^{12}$CO(1--0), $^{12}$CO(2--1), and $^{12}$CO(3--2)),
within a radius of 80~pc, forming a circumnuclear molecular disk. 
A detailed analysis of the kinematics, however, does not reveal any evidence of fueling of the nucleus.

The panels of the fourth line of Fig.~\ref{fig:overlays} show the comparison 
between \sigmahtwo\ and \sigmasfr\ distributions in NGC~4826 at a resolution of 
$\sim$2\farcs1 ($\sim$56~pc from $^{12}$CO(1--0)), $\sim$0\farcs7 ($\sim$17~pc from $^{12}$CO(2--1)),
and $\sim$2\farcs3 ($\sim$59~pc from $^{12}$CO(3--2)).
Unlike NGC~4569 and NGC~4579, the morphologies of \sigmasfr\ and \sigmahtwo\ in
NGC~4826 are spatially coincident and characterized by a structured disk. 

Figure \ref{fig:ks-n4826} and the righthand panel of Fig.~\ref{fig:co32}
show the molecular K-S relation derived for NGC~4826 with \sigmahtwo\ estimated from the three lowest 
$^{12}$CO emission lines at the intrinsic resolution of the gas maps.
As in NGC~3627, there are some cases where only four or five data points are involved 
in the analysis ($^{12}$CO(2--1) and $^{12}$CO(3--2) at 200~pc).
Neglecting these cases, the K-S relation derived from $^{12}$CO(1--0) shows 
that $N_{\rm fit}$ decreases from $\sim$$1.3$ at 200~pc-resolution to $\sim$$0.9$ at 56~pc-resolution.
The case with $^{12}$CO(1--0) at 200~pc has a superlinear $N_{\rm fit}$, 
while those analyzed at the intrinsic resolution of the all three $^{12}$CO maps (17--59~pc) are sublinear.
The comparison between the K-S relation with $^{12}$CO(1--0) at 56~pc and that with $^{12}$CO(3--2) at 59~pc
gives an almost identical $N_{\rm fit}$ ($\sim$0.9).

The good agreement between \sigmasfr\ and \sigmahtwo\ distributions (Fig.~\ref{fig:overlays}) 
consequently corresponds to a high Pearson correlation coefficient, equal to $\sim$0.7--0.8 for all the cases with more than 
five data points at resolutions of 56--200~pc.
These values of $r_{\rm corr}$ are the best in our sample of galaxies, and those of
$r_{\rm corr}$ drop down to $\sim$0.5 in the case using the $^{12}$CO(2--1) 
line at 17~pc, the highest resolution available not only in NGC~4826 but also in the present study.
This drop of $r_{\rm corr}$ could be due to the high resolution;
however, the comparison with the K-S relationship derived using the $^{12}$CO(2--1) line at 200~pc 
is not definitive proof of this because of the poor number statistics.
For other sample galaxies (NGC~4569 and NGC~4579), there is no a strong difference in $r_{\rm corr}$ between 
$^{12}$CO(1--0) and $^{12}$CO(2--1).
In addition to $N_{\rm fit}$, $r_{\rm corr}$  also assumes similar values (0.7--0.8) in the cases of 
$^{12}$CO(1--0) at 56~pc and $^{12}$CO(3--2) at 59~pc.
The results obtained for $N_{\rm fit}$ and $r_{\rm corr}$ therefore
suggest that NGC~4826 has a K-S relationship independent 
of the $^{12}$CO transition at resolutions as high as $\sim$60~pc.

Like NGC~4579, NGC~4826 has a \taudepl\ of $\sim$2~Gyr without radial trend both by using $^{12}$CO(1--0) 
and $^{12}$CO(2--1), when taking only the statistically significant cases into account.
The use of the $^{12}$CO(3--2) line instead gives a shorter \taudepl, of $\sim$1~Gyr, and radial trend
since it goes from $\sim$1~Gyr within $r < 1.0$~kpc to $\sim$0.7~Gyr within $r < 0.2$~kpc.

\begin{sidewaystable*}
\caption[]{Derived parameters from the pixel-by-pixel molecular star-formation relation for each galaxy.}
\begin{center}
\tiny
\begin{tabular}{ccclcccccccccccc}
\hline
\hline
Galaxy & $^{12}$CO transition & Spatial resolution & FoV$_{\rm sky}$ & $r_{\rm gal}$ & $N_{\rm fit(OLS\,bis.)}$ & $A_{\rm fit(OLS\,bis.)}$ 
& $r_{\rm corr(OLS\,bis.)}$ (n. pts)  & $N_{\rm fit(RR)}$ & $A_{\rm fit(RR)}$ & <$\Sigma{\rm_{H2}}$> & <$\Sigma{\rm_{SFR}}$> & \taudepl & scale \\
        &                      & [pc]              & [$^{\prime\prime}$] & [kpc] &     &     &  & &     & [M$_{\odot}$\,pc$^{-2}$] & [M$_{\odot}$\,yr$^{-1}$\,kpc$^{-2}$]& [Gyr] & [\arcsec/pix]\\
  (1)   & (2)                  & (3)                & (4)                 & (5)   & (6) & (7) & (8)                       & (9) & (10) & (11) & (12) & (13) & (14) \\
\hline
NGC 3627    & (1--0) & 82   & 24 & 1.3 & $1.18 \pm 0.11$ & $-3.66 \pm 0.19$ & 0.69 (12) & $ 0.83 \pm 0.27$ & $-2.72 \pm 0.73$ & 50.69 & 0.04 & 1.17 & 2.0 \\
                     & (1--0) & 200  & 24 & 1.3 & $1.11 \pm 0.41$ & $-3.47 \pm 0.74$ & 0.70 (4)  & $0.75 \pm 0.18$  & $-2.47 \pm 0.47$ & 53.60 & 0.04 & 1.49 & 4.0 \\ 
                     & (2--1) & 36   & 20 & 1.1 & $1.16 \pm 0.05$ & $-3.70 \pm 0.07$ & 0.62 (88) & $0.72 \pm 0.10$ & $-2.63 \pm 0.27$ & 82.32 & 0.06 & 1.27 & 0.8 \\
                     & (2--1) & 200  & 20 & 1.1 & $1.59 \pm 0.63$ & $-4.76 \pm 0.92$ & 0.65 (4)  & $1.06 \pm1.05$  & $-3.32 \pm 2.79$ & 75.19 & 0.05 & 1.45 & 4.0 \\
\hline
NGC 4569    & (1--0) & 154 & 30 & 3.1 & $0.60 \pm 0.03$ & $-1.96 \pm 0.03$ & 0.42 (37)  & $0.18 \pm 0.08$ & $-1.02 \pm 0.19$  & 51.40   & 0.16 & 1.03 & 2.0 \\   
            & (1--0) & 200 & 30 & 3.1 & $0.61 \pm 0.05$ & $-2.02 \pm 0.06$ & 0.43 (30)  & $0.24 \pm 0.10$ & $-1.21 \pm 0.22$ & 50.43   & 0.21 & 0.99 & 2.5 \\  
            & (2--1) &  74 & 20 & 2.1 & $0.59 \pm 0.01$ & $-1.99 \pm 0.01$ & 0.59 (145)  & $0.32 \pm 0.04$ & $-1.37 \pm 0.09$ &  116.29  & 0.36 & 1.17 & 1.0 \\
            & (2--1) & 200 & 20 & 2.1 & $0.64 \pm 0.05$ & $-2.11 \pm 0.11$ & 0.54 (20)   & $0.35  \pm 0.13$  & $-1.45 \pm 0.30$ & 92.68   & 0.31 & 1.12 & 2.5 \\
            & (3--2) & 190 & 30 & 2.9 & $0.67 \pm 0.04$ & $-2.23 \pm 0.11$ & 0.27 (39)   & $0.07 \pm 0.06$ & $-0.96 \pm 0.14$ & 46.50   & 0.27 & 1.05 & 2.5 \\
            & (3--2) & 200 & 30 & 2.9 & $0.69 \pm 0.08$ & $-2.28 \pm 0.08$ & 0.30 (36)   & $0.08 \pm 0.07$  & $-0.99 \pm 0.16$  & 42.63   & 0.25 & 1.03 & 2.5 \\
\hline
NGC 4579    & (1--0) & 154 & 36 & 2.7 & $0.80 \pm 0.04$ & $-2.94 \pm 0.02$ & 0.60 (65)  & $0.47 \pm  0.08$ & $-2.27 \pm 0.16$ & 17.36 & 0.01 & 2.28 & 1.8 \\
            & (1--0) & 200 & 36 & 2.7 & $0.91 \pm 0.01$ & $-3.13 \pm 0.02$ & 0.58 (61) & $0.57 \pm 0.09$ & $-2.48 \pm 0.18$ & 17.05 & 0.01 & 2.07 & 2.0 \\  
            & (2--1) & 76  & 20 & 1.5 & $0.54 \pm 0.01$ & $-2.29 \pm 0.02$ & 0.49 (111) & $0.24 \pm 0.04$ & $-1.69 \pm 0.08$ & 40.64 & 0.02 & 2.22 & 1.0 \\    
            & (2--1) & 200 & 20 & 1.5 & $1.14 \pm 0.03$ & $-3.49 \pm 0.03$ & 0.62 (40)  & $0.71 \pm 0.16$ & $-2.66 \pm 0.31$ & 37.76 & 0.02 & 1.60 & 2.0\\
\hline            
NGC 4826   & (1--0) & 56  & 40                    & 1.1 & $0.87 \pm 0.03$ & $-2.94 \pm 0.03$  & 0.72 (70)  & $0.64 \pm 0.07$ & $-2.43 \pm 0.17$ & 41.38  & 0.02  & 1.77 & 2.0 \\ 
           & (1--0) & 200 & 40$^{\mathrm{(a)}}$   & 1.1 & $1.27 \pm 0.08$ & $-3.89 \pm 0.05$  & 0.77 (12)  & $0.91 \pm 0.25$ & $-3.09 \pm 0.53$ & 105.07 & 0.05  & 1.98 & 7.6 \\
           & (2--1) & 17  & 20                    & 0.6 & $0.67 \pm 0.01$ & $-2.45 \pm 0.01$  & 0.52 (158) & $0.33 \pm 0.04$ & $-1.71 \pm 0.10$ & 112.31 & 0.05  & 2.07 & 1.0 \\    
           & (2--1) & 200 & 20$^{\mathrm{(b)}}$   & 0.6 & $1.64 \pm 1.31$ & $-4.17 \pm 0.86$  & 0.49 (4)   & $0.90 \pm 1.13$ & $-2.71 \pm 2.26$ & 99.85  & 0.13  & 0.79 & 7.6 \\   
           & (3--2) & 59  & 36                    & 1.0 & $0.89 \pm 0.04$ & $-2.73 \pm 0.07$  & 0.76 (27)  & $0.64 \pm 0.11$ & $-2.19 \pm 0.24$ &  16.28 & 0.02  & 1.07 & 2.0 \\  
           & (3--2) & 200 & 36$^{\mathrm{(c)}}$   & 1.0 & $1.22 \pm 0.11$ & $-3.19 \pm 0.17$  & 0.91 (5)  & $1.11 \pm 0.30$ & $-3.00 \pm 0.53$ &  20.93 & 0.03  & 0.61 & 7.6 \\      
\hline
\hline
\end{tabular}
\label{tab:fits}
\end{center}
\tablefoot{
\tablefoottext{a}{The effective FoV is reduced to 30\farcs4 since we rebinned 40$^{\prime\prime}$ with a pixel scale of 7\farcs6.}
\tablefoottext{b}{The effective FoV is reduced to 15\farcs2 since we rebinned 20$^{\prime\prime}$ with a pixel scale of 7\farcs6.}
\tablefoottext{c}{The effective FoV is reduced to 30\farcs4 since we rebinned 36$^{\prime\prime}$ with a pixel scale of 7\farcs6.}
}
\end{sidewaystable*}

\begin{figure*}
\begin{minipage}[b]{18cm}
\centering
\includegraphics[width=0.45\textwidth]{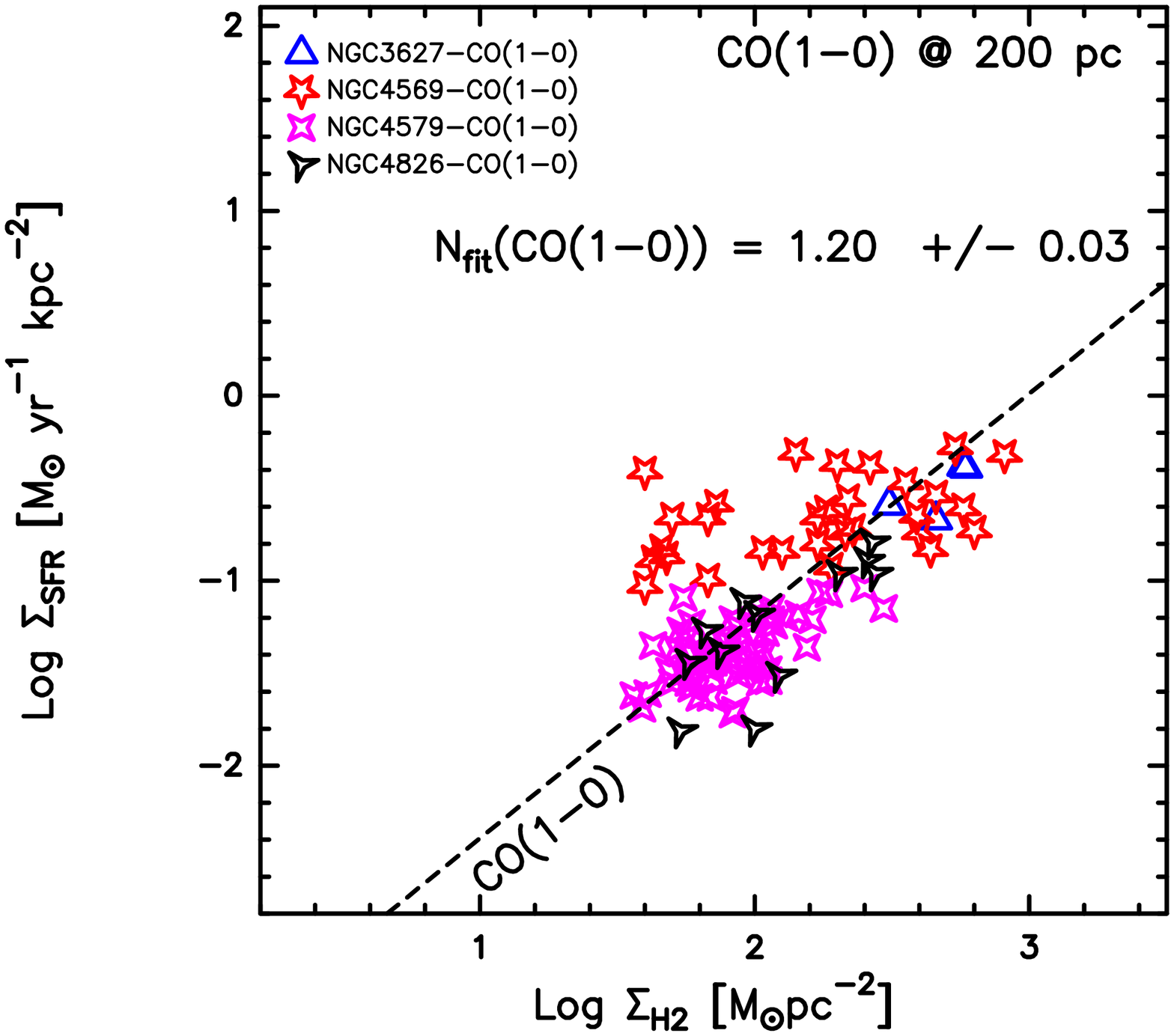}
\hspace{1cm}
\includegraphics[width=0.45\textwidth]{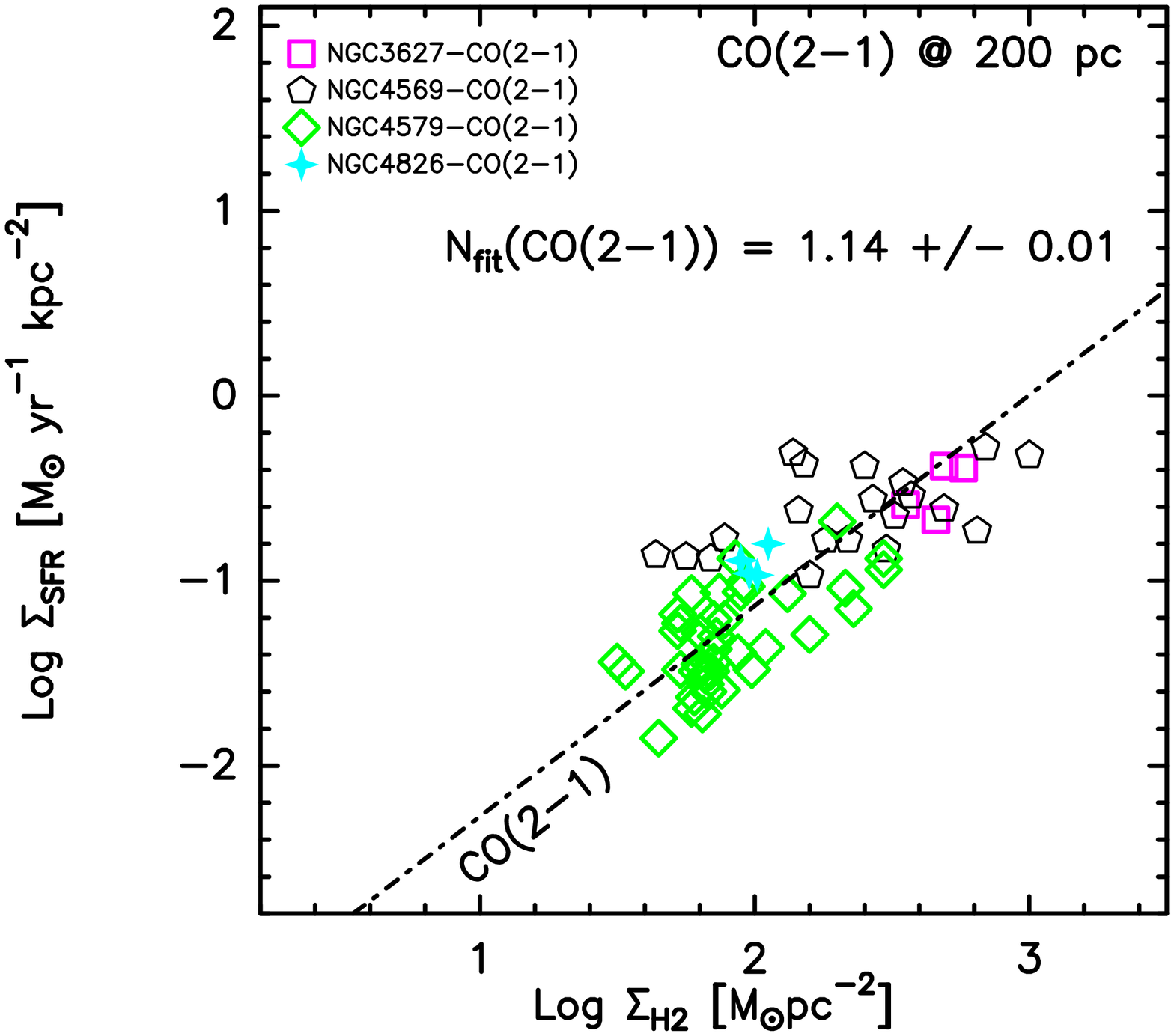}
\includegraphics[width=0.45\textwidth]{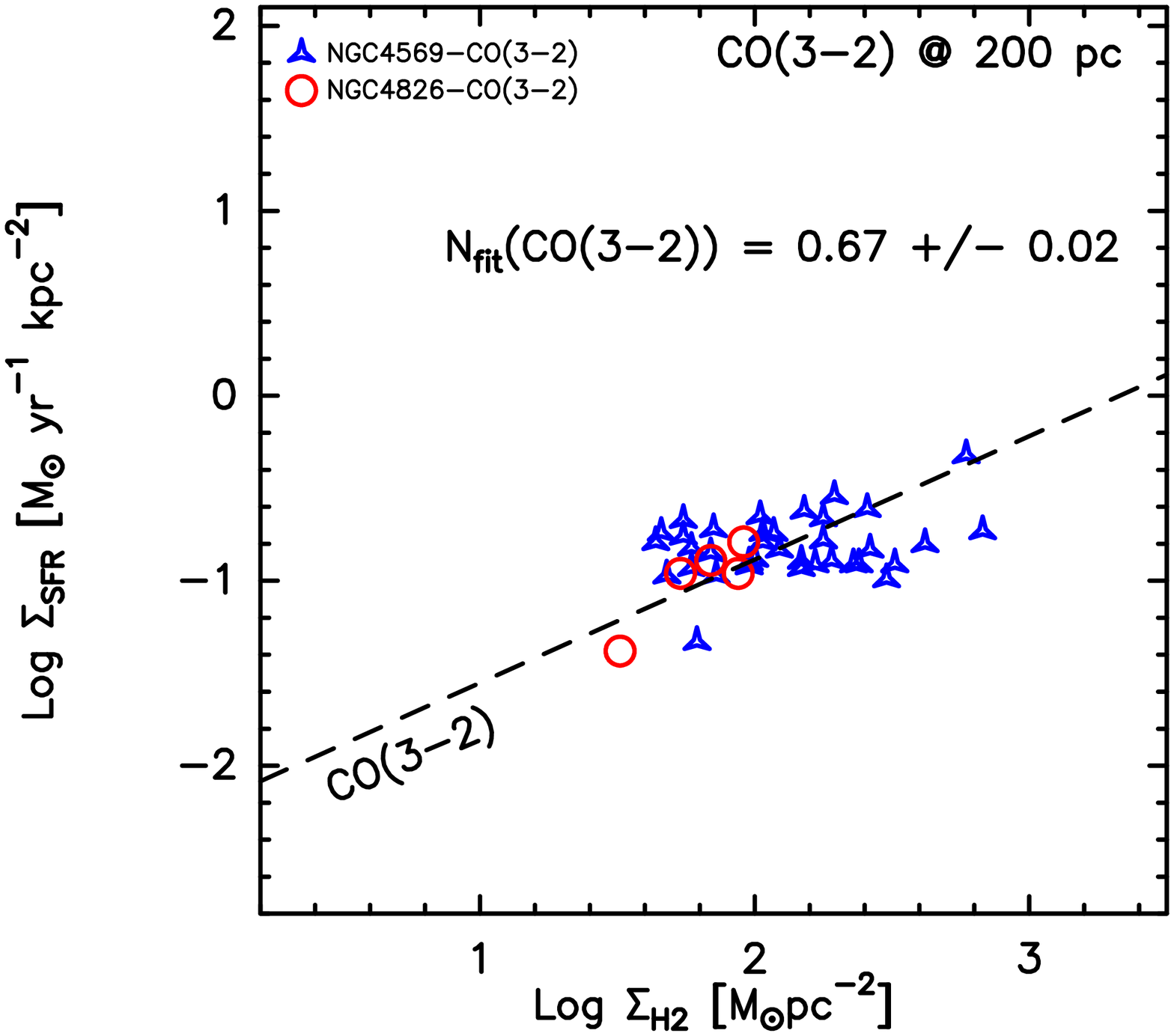}
\hspace{1cm}
\includegraphics[width=0.45\textwidth]{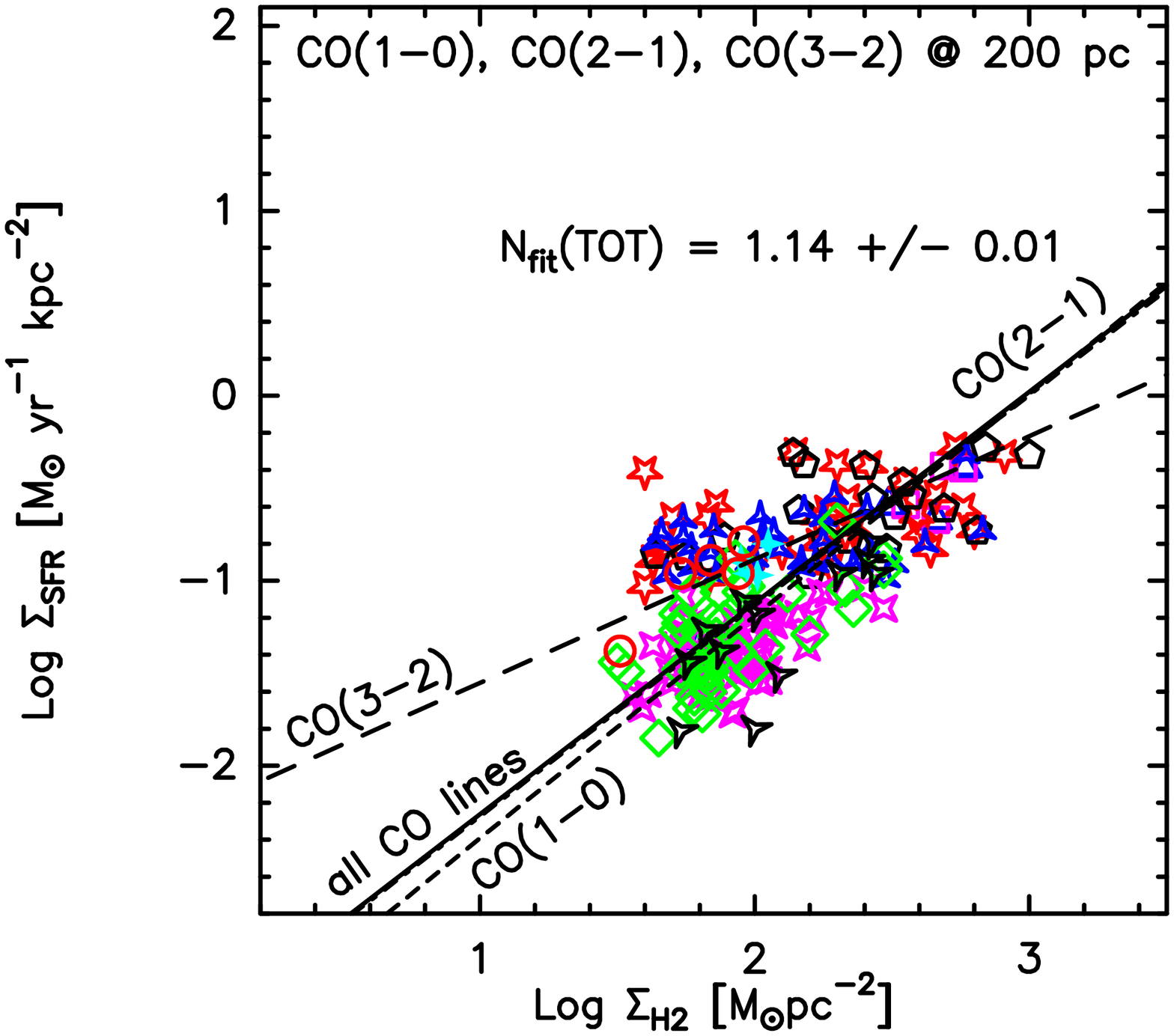}
\end{minipage}
\caption{
K-S relation plot for all sample galaxies.
The top left panel plots all galaxies and only $^{12}$CO(1--0) emission line data; 
the top right panel all galaxies and only $^{12}$CO(2--1) emission line data; 
the bottom left panel all galaxies and only $^{12}$CO(3--2) emission line data; and 
the bottom right panel all galaxies and all the available $^{12}$CO emission line data, 
at the common spatial resolution of 200~pc.
Different symbols and/or colors indicate galaxies whose \sigmahtwo\ 
has been derived from a given $^{12}$CO transition, as shown in the labels.
The black lines refer to the OLS fits derived for different cases:
the short-dotted line is for all galaxies and $^{12}$CO(1--0) emission line data,
the point-dotted line for all galaxies and $^{12}$CO(2--1), the long-dotted line for all galaxies and $^{12}$CO(3--2), 
and the black solid line is for all galaxies and all the available $^{12}$CO lines.
In the bottom right panel the $^{12}$CO(2--1) OLS fit line is hidden by total OLS fit line.
\label{fig:all-gal}
}
\end{figure*}

\section{The molecular star-formation relation across the sample}
\label{sec:sf-sample}
The analysis performed for each galaxy showed a wide range of behaviors in terms of the K-S relations,
and there is no ``universal'' molecular SF relation
although the investigated galaxies belong to the same subclass of objects (i.e., nearby active galaxies), 
and all the derived quantities have been treated with the same methodology and at comparable spatial resolutions.
The main result is therefore that each galaxy has its own SF relation (with its own index $N_{\rm fit}$, 
correlation coefficient, and \taudepl) on spatial scales of $\sim$20--200~pc. 
Nevertheless, we identified some common behaviors in terms of K-S relation, as discussed below.

\begin{table}
\caption[]{Derived parameters from the pixel-by-pixel molecular star-formation relation across the sample at the
common spatial resolution of 200~pc.}
\begin{center}
\begin{tabular}{cccc}
\hline
\hline
$^{12}$CO transition       & $N_{\rm fit}$             & $A_{\rm fit}$    & $r_{\rm corr}$ (n.pts) \\
          (1)        & (2)             & (3)              & (4)                   \\
\hline
(1--0)               & $1.20 \pm 0.03$ & $-3.60 \pm 0.01$ &  0.66 (107) \\
\hline
(2--1)               & $1.14 \pm 0.01$ & $-3.41 \pm 0.02$ & 0.77 (68) \\
\hline
(3--2)               & $0.67 \pm 0.02$ & $-2.22 \pm 0.08$ &  0.41 (41)\\
\hline
(1--0)+(2--1)+(3--2) & $1.14 \pm 0.01$ & $-3.41 \pm  0.01 $  & 0.64 (216)  \\
\hline
\hline
\end{tabular}
\label{tab:results-all}
\end{center}
\end{table}

\subsection{The K-S relation index: resolution vs. $^{12}$CO transitions}
\label{sec:sf-index}
By using the three lowest $^{12}$CO at resolutions of $\sim$20--200~pc,  
we found K-S relation indexes $N_{\rm fit}$ ranging from $\sim$0.5 to $\sim$1.6, which are all values that are consistent with literature results.
A superlinear slope of the K-S relation is consistent both with the early 
\text{global disk} 
studies
of the SF relation by \citet{kennicutt98a} and \citet{kennicutt98b} based on the combination of atomic and molecular gas data for the 
$\Sigma_{\rm gas}$ computation and with more recent works based only on molecular component at sub-kpc scales 
\citep[e.g.,][]{kennicutt07,verley10,liu11,rahman11,momose13}.
\citet{bigiel08} instead derived $N \approx 1$ from the correlation between \sigmasfr\ and \sigmahtwo\ estimated 
from $^{12}$CO(2--1) data for seven nearby spiral galaxies.
As already mentioned in Sect.~\ref{sec:intro}, \citet{bigiel08} suggest that a linear correlation is evident in regions 
of high gas surface densities where the gas is typically 
molecular ($\gtrsim$10~M$_{\odot}$\,pc$^{-2}$). 
Other studies using $^{12}$CO(2--1) also show a linear correlation \citep[e.g.,][]{leroy08,schruba11},
including one that combined single-dish $^{12}$CO(2--1) data with interferometric $^{12}$CO(1--0) data
\citep[][]{rahman11}.
These $^{12}$CO(2--1) studies analyzed a substantial number of nearby galaxies, though it should 
be recognized that some studies based on $^{12}$CO(1--0) data showed a superlinear (power-law) correlation, 
rather than a linear correlation \citep[e.g.,][]{wong02,kennicutt07,liu11}. 
This suggests that the choice of the CO transition for deriving \sigmahtwo\ could affect
$N_{\rm fit}$ values \citep[see][]{bigiel08}.
A linear slope has also been found by \citet{vutisalchavakul14} for the SF relation in an
11~deg$^2$ region of the Galactic plane with dust continua at 1.1~mm and 22~$\mu$m emission 
used as tracers of molecular gas and SFR, respectively, over a range of resolution from 33$^{\prime\prime}$
to 20$^{\prime}$ ($\sim$0.1--45~pc).

The most common explanation for a linear K-S relationship is that the observed CO luminosity 
is directly proportional to the number of star-forming clouds or GMCs, with all clouds
having similar properties, such as the volume density, the efficiency of the cloud, and the SFR.
In observations at resolution $\ga$100~pc, the individual clouds are not resolved but rather 
their CO flux is dispersed throughout the beam. 
In this case, regions with more clouds emit more CO in proportion to the number of clouds. 
Other recent works favor a sublinear ($N \sim 0.6 - 0.8$) K-S relationship at resolutions $\gtrsim$170~pc
\citep[e.g.,][]{blanc09,ford13,shetty13,shetty14a}.
A sublinear K-S relationship, in contrast to a linear one, suggests that the clouds do not have the same properties, 
and SFRs and/or volume densities vary \citep{shetty14b}.
This means that there is no one-to-one correspondence between the CO luminosity and the number of clouds. 
In addition, the conversion factor $X_{\rm CO}$ also varies with the location in a galaxy
(see later Sect.~\ref{sec:derivation-sh2}).
Finally, a more straightforward explanation for the sublinear K-S relationship would be that CO permeates
the hierarchical interstellar medium, including the filaments and lower density regions within
which GMCs are embedded \citep{shetty13,shetty14b}.

>From Table~\ref{tab:fits} it emerges that for a given $^{12}$CO transition, the index $N_{\rm fit}$ tends to gradually 
decrease with finer resolution down to a resolution of approximately 20~pc.
On the other hand, on scales larger than $\sim$100~pc, the slope of the SF relation tends to
increase and become somewhat superlinear. 
This result is common to all three lowest $^{12}$CO transitions and to all galaxies except for NGC~4569 
whose $N_{\rm fit}$ is almost constant as a function of resolution as shown in the previous section.
Our study therefore suggests that $\sim$80--100~pc is the scale at which the K-S relation undergoes a change, 
both with respect to kpc scales and slightly finer spatial scales
\citep[$\lesssim$250~pc, e.g.,][and this work]{bigiel08}. 
This contrasts with previous results 
by \citet{onodera10} for the Local Group spiral galaxy M~33, in which the slope steepens,
and the correlation virtually disappears on small spatial scales of
$\sim$80~pc (i.e., GMC scales). 
However, in M~33, both the SFR and gas surface densities at 80~pc resolution are significantly
lower than in the NUGA sample.

For distinguishing effects on the K-S relation index of different $^{12}$CO transitions 
to derive \sigmahtwo, data are needed at a common resolution, 
200~pc in the present work.
The results from this comparison are displayed and collected in Fig.~\ref{fig:all-gal} and
Table~\ref{tab:results-all}, respectively.
Each  panel of Fig.~\ref{fig:all-gal} shows the K-S relation plot for all sample galaxies and taking the three $^{12}$CO lines into account 
both separately and all together without distinction
based on the transition.
Different symbols and/or colors indicate galaxies whose \sigmahtwo\ 
has been derived from a given $^{12}$CO transition, and the black lines refer to the OLS fits 
derived for different cases.
At the resolution of 200~pc, taking all sample galaxies into account and $^{12}$CO(1--0) 
(top left panel) and $^{12}$CO(2--1) (top right panel)
separately, the resulting K-S relations 
have indices of $1.20 \pm 0.03$ and $1.14 \pm 0.01$, respectively.
These findings indicate that $^{12}$CO(1--0) and $^{12}$CO(2--1) lines lead to quite similar results in terms of 
the slope of the K-S relation.
Under the same conditions but by using $^{12}$CO(3--2) data (bottom left panel), we found a sublinear slope 
of $0.67 \pm 0.02$.
However, this result is based only on two cases, NGC~4569 and NGC~4826, and for NGC~4569
all slopes for the three lowest $^{12}$CO transitions at all investigated resolutions (74--200~pc) are sublinear. 
In contrast, in NGC~4826 the slope derived from the $^{12}$CO(3--2) line is roughly unity at 200~pc
of resolution (see Table~\ref{tab:fits}).
Taking all sample galaxies and all available $^{12}$CO transitions (bottom left panel) into account, 
the K-S relation has a ``standard'' index of $1.14 \pm 0.01$.

It is well known that each transition traces different physical gas properties.
The kinematic temperature of molecular gas is typically $\sim$10~K \citep[][]{scoville87}, which is above the level energy temperature 
of 5.5~K for the $J = 1$ level of the $^{12}$CO, but below the temperatures of 16.5~K and 33~K for the $J = 2$ and 
$J = 3$ levels, respectively.
This implies that a slight change in gas kinematic temperature is sufficient to affect the excitation for the $^{12}$CO(2--1) 
and $^{12}$CO(3--2) emission lines.
The different critical densities of the three $^{12}$CO transitions ($\sim$10$^{3}$~cm$^{-3}$, $\sim$$2 \times 10^4$~cm$^{-3}$,
and $\sim$$7 \times 10^4$~cm$^{-3}$ for $^{12}$CO(1--0), $^{12}$CO(2--1), and $^{12}$CO(3--2), respectively) 
make their line ratios sensitive to local gas density.
 The value of $R_{21}$ has been observed to systematically vary with SFR and $\Sigma_{\rm gas}$ in the Milky Way 
\citep[e.g.,][]{sakamoto95,sawada01} and M~51 \citep[][]{koda12,vlahakis13}.
Despite this, because of the use of individually measured mean $R_{\rm 32}$ to convert to an equivalent CO(1--0) mass
conversion, such variations do not have a significant 
impact on the resulting index $N_{\rm fit}$ of the K-S relation.

Another result emerging from Table~\ref{tab:fits} concerns the K-S relation studied 
at 200~pc resolution with \sigmahtwo\ derived from $^{12}$CO(1--0).
NGC~3627, NGC~4579, and NGC~4826 exhibit K-S relations with indices $N_{\rm fit}$ around unity, 
with \sigmahtwo\ of $\sim$20--100~\msunpc, and \taudepl\ of $\sim$1.5--2~Gyr.
NGC~4569 is different from the other sample galaxies also in this respect; 
it indeed has a \sigmahtwo\ of $\sim$50~\msunpc, comparable to the other galaxies, 
but its K-S relation has a sublinear slope and \taudepl\ of $\sim$1~Gyr.

\begin{table*}
\caption[]{Stellar and star formation properties of our galaxy sample.}
\begin{center}
\begin{tabular}{ccccc}
\hline
\hline
Galaxy       & log(M$_\ast$) $^{\mathrm{(a)}}$ & log(TIR) $^{\mathrm{(b)}}$ & SFR $^{\mathrm{(c)}}$ & log(sSFR) \\
             & [M$_\odot$]                     & [erg~s$^{-1}$]             & [M$_\odot$~yr$^{-1}$]  & [yr$^{-1}$] \\
  (1)        & (2)                             & (3)                        & (4)                   & (5) \\
\hline
NGC~3627     & $10.57 \pm 0.13$ & 10.4 & 4.33 & $-9.93$ \\
\hline
NGC~4569     & $10.38 \pm 0.12$ & 9.7  & 0.86 & $-10.44$ \\
\hline
NGC~4579     & $9.96 \pm 0.23$  & 10.1 & 2.17 & $-9.62$ \\
\hline
NGC~4826     & $9.99 \pm 0.12$  & 9.6  & 0.69 & $-9.30$  \\
\hline
\hline
\end{tabular}
\label{tab:ssfr}
\end{center}
\tablefoot{
\tablefoottext{a}{Values from \citet{skibba11}.}
\tablefoottext{b}{Values from \citet{dale12}.}
\tablefoottext{c}{SFR derived adopting calibration from \citet{kennicutt98a} for TIR luminosity.}
}
\end{table*}

\subsection{The molecular gas depletion time}
\label{sec:depletion}
As discussed in Sect.~\ref{sec:intro}, \taudepl\ is an important parameter in the study of the K-S relation,
as is the power index $N_{\rm fit}$, for understanding the SF phenomenon in galaxies. 
We found molecular \taudepl\ ranging from $\sim$1 to $\sim$2~Gyr, which is consistent with more recent results as discussed below.
Each galaxy has its own \taudepl\ that seems to be invariant with respect to the spatial scale
probed in a given galaxy, at resolutions from $\sim$20 to 200~pc.
The molecular \taudepl\ is also invariant with respect to $^{12}$CO(1--0) and $^{12}$CO(2--1) lines.
The discussion of the \taudepl\ derived from $^{12}$CO(3--2), for NGC~4569 and NGC~4826, deserves separate treatment and is discussed later.

Both NGC~3627 and NGC~4569 have molecular \taudepl\  of $\sim$1--1.3~Gyr, 
by using $^{12}$CO(1--0) and $^{12}$CO(2--1) and taking only the statistically significant cases into account.
These values of \taudepl\ are compatible with the mean molecular [$^{12}$CO(1--0)] \taudepl\ of 
$\sim$1.2~Gyr (with $\alpha_{CO} = 3.5$~$M_{\odot}$~(K~km~s$^{-1}$~pc$^{-2}$)$^{-1}$) found by \citet{saintonge11} 
for the COLD GASS sample of more than 200 galaxies at distances of $\sim$100--200~Mpc,
but lower than the mean molecular [$^{12}$CO(2--1)] \taudepl\ of $\sim$2~Gyr found by \citet{bigiel08} and \citet{leroy08} in 
the HERACLES survey consisting of a sample of 48 nearby ($D \sim 3 - 20$~Mpc) galaxies.
NGC~3627 and NGC~4569 therefore have a \taudepl\ that is more consistent with the COLD GASS single-dish survey that gives a global picture of gas 
and SF in the Local Universe but lacks the power to trace the exact distribution of these components, 
rather than with the HERACLES survey, which like the present study, maps them. 
However, \citet{saintonge11} demonstrate that after using the same conversion factor from CO luminosity and H$_2$ mass
($\alpha_{CO} = 3.5$~$M_{\odot}$~(K~km~s$^{-1}$~pc$^{-2}$)$^{-1}$) 
and restricting the HERACLES sample to the COLD GASS stellar mass range, the mean molecular \taudepl\ 
for HERACLES is $\sim$1~Gyr, which is consistent with the COLD GASS estimate.

As mentioned in Sect.~\ref{n4569}, \taudepl\ of NGC~4569 also shows a hint of radial trend-reaching values of $\sim$0.7--0.9~Gyr within a radius of 0.5~kpc.
These low values for the molecular \taudepl\ are also in line with the recent results by \citet{faesi14} 
for NGC~300 on 250~pc scales, so slightly larger than but comparable to ours, and by \citet{lada10} for the Milky Way GMCs, 
who found depletion times of $\sim$0.2~Gyr.
After \citet{faesi14} studied the relation between molecular gas and SF in 76 H{\sc ii} regions of NGC~300,
they concluded that the short \taudepl\ arises because their analysis accounts for only 
the gas and stars within the youngest star-forming regions. 
These depletion times correspond to the timescale for SF to consume the gas reservoir in the star 
clusters' parent GMCs, which may be the more relevant quantity in the context of GMC-regulated SF 
in galaxies.

The hint of a radial trend of \taudepl\ in NGC~4569 also suggests that SF becomes more 
inefficient at larger radii under the assumption of a fixed $X_{\rm CO}$ conversion factor. 
This finding agrees with results obtained by \citet{leroy13} who find systematically lower \taudepl\ 
for a given $X_{\rm CO}$ in the inner kpc of a sample of 30 galaxies, among AGN and starbursts.
Since $X_{\rm CO}$ is typically smaller when closer to galaxy nuclei \citep[][]{sandstrom13}, implying  
shorter \taudepl\ for a given CO luminosity, the radial trend in \taudepl\ could be even steeper. 
The shortening of \taudepl\ toward the very center of NGC~4569 coincides both 
with a better correlation between \sigmahtwo\ and \sigmasfr\ distributions (see Fig.~\ref{fig:overlays}) 
and with an increase in the $R_{\rm 21}$ ratio there, which is indicative of more excited gas.
The value of $R_{\rm 21}$ ranges from 0.2 to 0.9 in the inner region of NGC~4569 \citep{boone07,boone11}, with a smooth and 
fairly symmetric distribution with respect to the center, and continuously increasing toward the nucleus.

Similar radial trends in $R_{\rm 21}$that are also present in other NUGA galaxies 
\citep[NGC~4579, NGC~6574:][respectively]{santi09,lindt-krieg08}
and in the sample of \citet{leroy13} presumably reflect
similar changing physical conditions that drive the lower $X_{\rm CO}$ factors found by 
\citet{sandstrom13}.
This would underscore that molecular gas in the central regions gives off more 
CO emission, appears more excited, and forms stars more rapidly than molecular gas 
farther out in the disks of galaxies. 
Compared to $R_{\rm 21}$, $R_{\rm 32}$ shows a clumpier and more asymmetric distribution within the central region
\citep{boone11}, differently from the large-scale gradient in $R_{\rm 32}$ reported by \citet{wilson09}.  Usually,
$R_{\rm 32}$ is expected to probe more extreme physical conditions, such as those 
occurring in star-forming or shock regions.
\citet{boone11} indeed confirm that $R_{\rm 32}$ tracks massive SF in NGC~4569 better than $R_{\rm 21}$
by comparing $R_{\rm 21}$ and $R_{\rm 32}$ maps to the Pa$\alpha$ image as 
detected by NICMOS \citep[see also][]{wilson09}.
In any case, in NGC~4569, $^{12}$CO is less excited ($R_{\rm 32} \sim 0.6$) than in the Galactic center and in most of the centers of 
normal, starburst, and active galaxies studied so far for which 
$R_{\rm 32} \approx 0.2-5$ \citep[e.g.,][]{devereux94,mauersberger99,matsushita04,mao10,francoise13,santi14}. 
Shorter \taudepl\ could also be due to the high pressure driven by the deep potential well 
in the central parts of galaxies, driving gas to higher densities.
This effect has been seen both in our Galaxy \citep{oka01} and in others \citep{rosolowsky05}, 
and it has been suggested that it explains shorter \taudepl\ (i.e., enhanced SFE 
in galaxy centers) in the sample of \citet{leroy13}.

NGC~4579 and NGC~4826 instead have higher molecular \taudepl\ of $\sim$2~Gyr,
both with $^{12}$CO(1--0) and $^{12}$CO(2--1) when only taking the statistically significant cases into account.
For NGC~4579, neglecting the case of the net sublinear K-S relationship 
found with $^{12}$CO(2--1) at the resolution of 76~pc, 
\taudepl~$\sim$2~Gyr is accompanied by an index $N_{\rm fit}$ around unity.
As already mentioned above, the interpretation of a linear K-S relationship is that CO
is primarily tracing star-forming clouds with relatively uniform properties, including \sigmasfr, and consequently
the \taudepl\ of the CO traced gas is constant and of $\sim$2~Gyr, both within and between galaxies.
However, there is still no consensus on either the precise K-S parameter estimates or the associated interpretation
\citep[see][for a review of explanations of the K-S relationship]{dobbs13}.
In contrast to NGC~4569, \taudepl\ in NGC~4579 does not show a radial trend although 
$R_{21}$ varies as a function of the radius.
Close to the AGN, at $r < 200$~pc, $R_{21}$~$\approx~$0.8--1, whereas at $r > 500$~pc $R_{21}$$\approx~$0.4--0.6
\citep[][]{santi09}.

The use of the $^{12}$CO(3--2) line gives a \taudepl\ of $\sim$1~Gyr both for NGC~4569 and NGC~4826, 
suggesting that the gas traced by this transition could be more efficiently converted into stars with respect 
to $^{12}$CO(1--0) and $^{12}$CO(2--1) emissions, despite using mean CO(1--0)/CO(3--2) ratios
to convert mass densities.
While NGC~4569 has \taudepl~$\sim$~1~Gyr for all three $^{12}$CO lines and for all resolutions taken into account, 
NGC~4826 shows a lower $^{12}$CO(3--2) \taudepl\ than those 
derived from $^{12}$CO(1--0) and $^{12}$CO(2--1) ($\sim$2~Gyr).
In addition, for NGC~4826,
this trend is seen by using $^{12}$CO(3--2), while \taudepl\ does not show a radial trend in
the cases of $^{12}$CO(1--0) and $^{12}$CO(2--1).
However, the $^{12}$CO(3--2) distribution is similar to that of the CO(1--0) and CO(2--1) lines at the
same resolution \citep[][]{boone11}.
In contrast to NGC~4569, $R_{32}$ does not appear to be clumpier than $R_{21}$ in NGC~4826,
and both maps look relatively symmetric with respect to the dynamical center, with both ratios increasing
steadily toward the center \citep[][]{boone11}.
As in NGC~4569, high $R_{32}$ values ($>$0.3) occur within the Pa$\alpha$ emission region as 
detected by NICMOS, while this is not true for $R_{21}$, which reaches high values ($>$0.7) away 
from the Pa$\alpha$ emitting regions \citep[][]{boone11}.
This difference, along with the variation in $R_{\rm 32}$, could be responsible
for the short \taudepl\ inferred by $^{12}$CO(3--2) in NGC~4826.

Compared to the $^{12}$CO(1--0) and $^{12}$CO(2--1), the $^{12}$CO(3--2) traces relatively warmer and denser gas. 
There is also growing evidence that the $^{12}$CO(3--2) emission correlates more tightly with 
the SFR or SFE than does $^{12}$CO(1--0) and $^{12}$CO(2--1) \citep[e.g.,][]{komugi07,muraoka07,wilson09,wilson12}.
Similarly, \citet{iono09} show that the $^{12}$CO(3--2) emission correlates nearly linearly 
with the FIR luminosity for a sample of local luminous IR galaxies and high-redshift submillimeter galaxies. 
Thus, it appears that the $^{12}$CO(3--2) emission primarily traces the molecular 
gas associated directly with SF, such as high-density gas that is forming stars or warm gas heated by SF, 
rather than the total molecular gas content of a galaxy usually traced by $^{12}$CO(1--0) and $^{12}$CO(2--1).
Nevertheless, this evidence is not able to explain the shorter \taudepl\ characterizing the $^{12}$CO(3--2)-molecular K-S 
relation for NGC~4826 with respect to $^{12}$CO(1--0) and $^{12}$CO(2--1).

\begin{figure}
\centering
\includegraphics[width=0.45\textwidth]{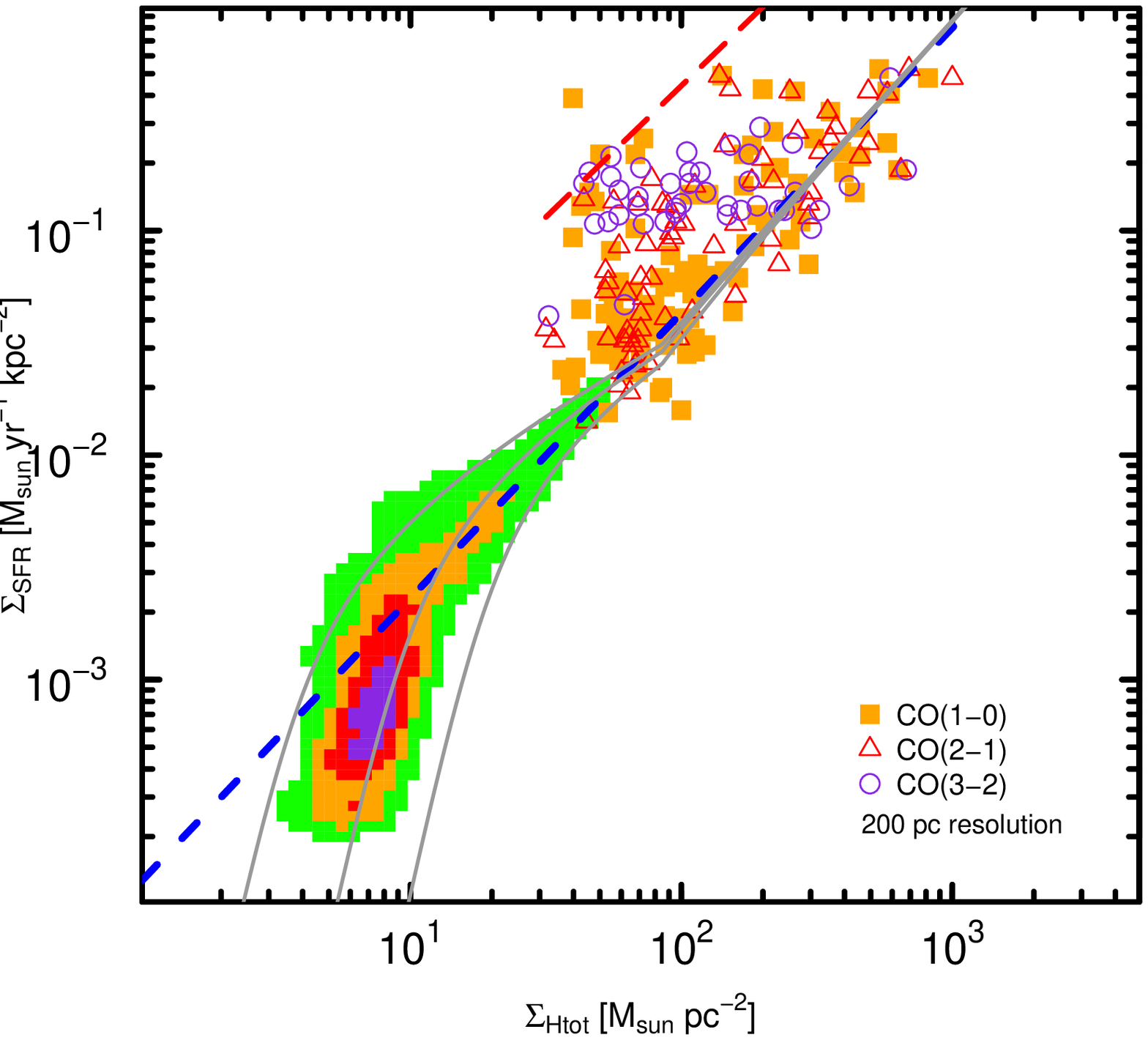}
\caption{
Comparison between (i) the K-S relation plot derived for all sample galaxies and taking all the $^{12}$CO emission lines  
at the common resolution of 200~pc into account, as shown in Fig.~\ref{fig:all-gal}, and (ii) results from \citet{bigiel08}.
The model predictions from \citet{krumholz09} are also shown with solar metallicity and
clumping factor $c$ = 1, 2, and 5; see \citet{krumholz09} for more details.
Also shown (as a blue short-dashed line) is the trend for star-forming galaxies and mergers found by 
\citet{genzel10} using a common conversion factor $\alpha_{\rm CO}$ ($= 3.2$~M$_{\odot}$~pc$^{-2}$~(K~km~s$^{-1}$)$^{-1}$). 
The horizontal axis has been adjusted to our $\alpha_{\rm CO}$ 
($= 3.5$~M$_{\odot}$~pc$^{-2}$~(K~km~s$^{-1}$)$^{-1}$) and our non-inclusion of helium.
We have also plotted the merger/starburst trend by \citet{genzel10} 
(with $\alpha_{\rm CO} =1$~M$_{\odot}$~pc$^{-2}$~(K~km~s$^{-1}$)$^{-1}$) as shown by a (red) long-dashed line. 
\label{fig:nuga-bigiel}
}
\end{figure}

Values of \taudepl\ derived for a given galaxy and within comparable FoVs 
are indeed expected to be similar for all $^{12}$CO lines, by construction, since 
we are not deriving gas traced by $^{12}$CO(3--2) but rather the total gas, assuming 
a $R_{\rm 32}$ line ratio, in addition to the $X_{\rm CO}$ conversion factor (see Eq.~\ref{s32}). 
The bottom righthand panels of Figs.~\ref{fig:all-gal} and \ref{fig:nuga-bigiel} 
show that \sigmahtwo\ data points derived from the $^{12}$CO(3--2) line 
emission have typically lower \sigmahtwo\ values with respect to $^{12}$CO(1--0) and $^{12}$CO(2--1) derivation, 
but this effect is not due to the choice of the $^{12}$CO line but simply to the availability of $^{12}$CO(3--2) data 
only for NGC~4569 and NGC~4826. 
It is possible that our result is simply due to the
variation in $R_{\rm 32}$ that could not be considered in the mean value used
for the derivation of \sigmahtwo\ in Eq.~\ref{s32}.

The radial trend of \taudepl\ as a function of physical conditions of gas, typically expressed in terms of line ratios,
is not clearly defined.
While NGC~4569 exhibits a \taudepl\ that decreases toward the very center, together with an increase in
$R_{\rm 21}$, NGC~4579 has constant \taudepl\ as a function of the radius but higher $R_{\rm 21}$
in the central galaxy region.      
This difference could be explained by the (dis)agreements between gas and SFR morphologies for the two galaxies.
Both NGC~4569 and NGC~4579 are centrally peaked in \sigmasfr\ without a counterpart in \sigmahtwo\ (see Fig.~\ref{fig:overlays}).
However, NGC~4569 shows CO emission in the very center, but NGC~4579 has neither gaseous nuclear emission 
nor a radial trend in the gas morphology (although the $R_{\rm 21}$ line ratio decreases as a function of radius).
It appears therefore that the radial trend of \taudepl\ as a function of CO line ratios also has to be accompanied by
similar radial trends in both gas and SFR morphologies in a given galaxy.

It is well known that gas depletion times are different in extreme starburst galaxies and merging systems
in the local Universe, as well as in submillimeter galaxies at high redshifts 
\citep[e.g.,][]{kennicutt98a,bouche07,riechers07,bothwell10}. 
In these galaxies, gas depletion times are significantly shorter than the normal galaxy population, with the
star formation surface density
being an order of magnitude larger at fixed gas surface density 
\citep[][]{genzel10}.
More recently, it has emerged that the molecular \taudepl\ varies even within the population of nearby 
`normal' star-forming galaxies and that this variation has been quantified as a function of a range of fundamental
parameters.   
\citet{saintonge11}, \citet{boselli14}, and \citet{bothwell14} all find that depletion timescales vary monotonically from 
$\sim$2~Gyr to $\sim$0.1~Gyr across two orders of magnitude in stellar mass.
This finding also persists when using a constant conversion factor and different conversion factor values \citep{bothwell14}.
\citet{saintonge11} also find that \taudepl\ correlates inversely with the specific SFR (SFR per unit stellar masses, sSFR).
Table~\ref{tab:ssfr} collects the main stellar and star formation properties of our galaxy sample.
In this table, Col. (1) gives the galaxy name, Col. (2) the logarithm of the stellar mass (M$_\ast$), Col. (3) the logarithm of the 
total IR (TIR) luminosity, Col. (4) the SFR derived adopting the conversion of \citet{kennicutt98a} for TIR (from 8--1000~$\mu$m),
and Col. (5) the logarithm of the sSFR.
NUGA galaxies have stellar masses in the COLD GAS range (log(M$_\ast$) from 10 to 11.5~M$_\odot$) and sSFR toward 
the high end of the COLD GAS sample (log(sSFR) from $-11.8$ to $-9.5$ yr$^{-1}$),
so are consistent both with our results for short depletion times and the correlation that \citet{saintonge11} 
find between \taudepl\ and sSFR.
Nevertheless,
the NUGA galaxies with the longest \taudepl\ (although only a factor of 2)
have the highest sSFR, although the values are within the scatter found by \citet{saintonge11}.

The dependence of \taudepl\ on the spatial scale has also been explored by other groups. 
\citet{schruba10} find a median \taudepl\ $\sim$1.1~Gyr for M~33 (using the conversion factor we use here) with no dependence on type of region targeted
on kpc scales, but below these scales \taudepl\ is a strong function of both the adopted scale 
and the type of region that is targeted \citep[see also][for \taudepl\ even down to 0.1~Gyr
in the central few hundred parsecs of local spirals]{komugi05}.
\citet{schruba10} derived very long \taudepl\ (up to 10~Gyr) using small apertures centered on CO peaks, 
but very short \taudepl\ (0.3~Gyr) with small apertures targeted toward H$\alpha$ peaks.
This may not be surprising, given the disparity in the bright H$\alpha$ and CO distributions in M~33
\citep[see Fig.~1 in][]{schruba10}, similar to the inconsistencies between $\Sigma_{\rm gas}$
and $\Sigma_{\rm SFR}$ morphologies characterizing some NUGA galaxies, such as NGC~4569 and NGC~4579 
(see Fig~\ref{fig:overlays}).
\citet{schruba10} find that the SF relation in M~33 observed on kpc scales
breaks down for aperture sizes of $\sim$300~pc.
These aperture sizes are larger than the breakdown scale of 80--100~pc found by \citet{onodera10} 
for the same galaxy.

\subsection{The dispersion of the data}
\label{sec:dispersion}
At the relatively high resolutions studied here, the Pearson correlation coefficient $r_{\rm corr}$ is generally low, and 
there is large scatter 
with values as low as $\sim$0.3. Angular resolutions of $\sim$1--2\arcsec\ correspond to $\sim$20--200~pc at the distances of the sample galaxies.
An inspection of the maps 
shown in Fig.~\ref{fig:overlays}
reveals that CO intensity peaks are offset from the H$\alpha$ peak, often by more than $\sim$1--2\arcsec.
In addition, there is the possibility that such small beams do not include both the star-forming region and
its counterpart molecular cloud, which is generally several tens of parsecs in size.
In general, star-forming regions do not coincide with the molecular clouds that give birth to them.
This is typically seen in spiral arms of galaxies, where the optical and molecular spiral structures are offset.
If only the molecular clouds are observed and not the star-forming regions which are physically coupled to them,
the gas density would be expected to show an excess over the expected correlation.

For a given galaxy, for a given $^{12}$CO transition, and with only the statistically significant cases, $r_{\rm corr}$ 
is quite stable (see NGC~4569) or it tends to decrease with resolution, although this tendency is weak (see NGC~4579 
in $^{12}$CO(2--1) and NGC~4826 in $^{12}$CO(1--0)).
The latter trend is consistent with the result of a larger scatter on smaller scales found by \citet{vutisalchavakul14} 
when studying the SF relation in the Galactic plane.
They found that the rank correlation coefficients between log(\sigmahtwo) and log(\sigmasfr) increase from $\sim$0.4 
to $\sim$0.8 in the resolution range from 33$^{\prime\prime}$ to 20$^{\prime}$.
\citet{vutisalchavakul14} also compared their results with some resolved extragalactic SF relations, available 
on a scale of about 200~pc \citep[for instance, M~33 from][]{schruba10}, finding the same trend of the dispersion as a function of resolution for 
extragalactic data and a smaller scatter in the depletion time distribution for the Galactic plane than for the extragalactic 
sources.

\subsection{The K-S relation at high densities of molecular gas}
\label{sec:high-sh2}
The range of $\Sigma_{\rm H_2}$ over which we studied the K-S relation is $\sim$20--120~\msunpc.
These densities are higher than those of 20--50~\msunpc\ studied by \citet{bigiel08} and other authors.
The circumnuclear regions examined here are subject to physical conditions that are different from the disk.
Besides denser gas tracers, such as HCN and HCO$^{+}$, observed in these regions 
\citep[e.g.,][]{krips11,francoise13,francoise14,santi14}, rotation curves
in the central kpc are different from disks, following roughly rigid-body rotation 
compared to nearly flat in the disks \citep[e.g.,][]{sofue03}.
This would lessen the effect of shear, which works on the clouds, and may enhance the efficiency of the SF 
arising from gravitational collapse. 
These effects could conspire to produce a systematic change in the K-S relation in the central regions of galaxies.
Figure~\ref{fig:nuga-bigiel} 
shows the comparison between our results for the whole sample at 200~pc resolution
(as also displayed in Fig.~\ref{fig:all-gal}) and the findings from \citet{bigiel08}.
Also plotted in Fig. \ref{fig:nuga-bigiel} are the predictions from the \citet{krumholz09} models
for solar metallicity and various clumping factors \citep[see][for more details]{krumholz09}. 
For the comparison we have converted the \sigmahtwo\ measurements in \citet{bigiel08} to the
conversion factor we use here, but have not adjusted the SFRs since the Kroupa IMF \citep[][]{kroupa01} 
used by them is similar to ours \citep[see][]{calzetti07}.
Figure \ref{fig:nuga-bigiel} also shows two trends found by \citet{genzel10}:
the (blue) short-dashed line gives the common trend for star-forming galaxies (SFGs)
and mergers/starburst using a single conversion factor ($\alpha_{\rm CO} =3.2$~M$_{\odot}$~pc$^{-2}$~(K~km~s$^{-1}$)$^{-1}$);
the (red) long-dashed line corresponds to the merger/starburst trend only 
(with $\alpha_{\rm CO} =1$~M$_{\odot}$~pc$^{-2}$~(K~km~s$^{-1}$)$^{-1}$).
Both regressions have been adjusted to our $\alpha_{\rm CO}$
(3.5~M$_{\odot}$~pc$^{-2}$~(K~km~s$^{-1}$)$^{-1}$) and to our not including helium.

We find that at densities as high as $\sim$10$^{3}$~M$_{\odot}$~pc$^{-2}$ there is
reasonable agreement with the predictions of \citet{krumholz09}. 
However, with a common conversion factor,
the central regions of the NUGA galaxies have higher
\sigmasfr\ for a given gas column than would be expected from the models.
They fill in the region between the mergers or high-$z$ starburst systems and 
the more quiescent SFGs studied by \citet{genzel10}, 
assuming that the former require a lower value of $\alpha_{\rm CO}$.
The apparently high \sigmahtwo\
would also be consistent with the results of \citet{sandstrom13}, who find that
$\alpha_{\rm CO}$ in centers of galaxies relative to their disks can be as much as ten times 
lower.
We could thus be slightly overestimating \sigmahtwo\ in our sample, although it is difficult to determine by exactly how
much. 
To assuage the offset between mergers or starbursts and SFGs, 
\citet{krumholz12} have devised a ``universal SF law'' that is based on
volumetric densities \citep[see also][]{genzel10}. 
They argue that such a formulation is particularly important for resolutions of 10--100~pc, 
similar to what is achieved by our observations. 
In addition, the galaxies studied here are roughly consistent with model predictions of
\citet{krumholz05} and \citet{krumholz09}.

It is difficult to consider the K-S relation for normal galaxies at even higher densities.
This would require that we observe with even better spatial resolution (possible with ALMA), but 
the GMCs would then be highly resolved.
Thus we may not be able to observe the molecular cloud and its counterpart star-forming region 
within one beam, for galaxies -- such as those studied here -- at distances of $\sim$5--20~Mpc (see Sect.~\ref{sec:dispersion}). 
It is possible that the K-S relation is fundamentally a correlation that is seen only when the ISM is averaged over a considerable area.
As already discussed in Sect.~\ref{sec:sf-index}, our work shows that the K-S relation undergoes a change at a spatial resolution of $\sim$80--100~pc
($\sim$GMC scale); i.e., its slope starts to decrease sharply with respect to higher scales.
The breakdown of the K-S relation could therefore occur at $\sim$80--100~pc,
as found by \citet{onodera10} for M~33, even if in NUGA dense nuclei a net breakdown is still not visible on scales
as high as $\sim$20~pc.

\section{Caveats}
\label{sec:caveats}
Here we examine possible caveats or systematic uncertainties associated
with our analysis and results presented in the previous sections.  
These can arise from physical causes connected to our simplified assumptions, from 
the methodology we used to derive the various quantities, and/or from the adopted fitting method of data.

\subsection{Derivation of \sigmahtwo}
\label{sec:derivation-sh2}
It is well known that the $X_{\rm CO}$ conversion factor is a source of systematic uncertainty in 
the estimate of \sigmahtwo.
By adopting a constant value for $X_{\rm CO}$,
we do not account for its changes with galaxy properties, such as physical gas conditions and 
metallicity \citep[e.g.,][]{maloney88,bolatto08,glover11,magrini11,bolatto13}.
Estimates of ``typical'' values of $X_{\rm CO}$ in the Milky Way and other spiral galaxies range
from $\sim$1.5 $\times$ 10$^{20}$ cm$^{-2}$ (K km s$^{-1}$)$^{-1}$ to
$\sim$4 $\times$ 10$^{20}$ cm$^{-2}$ (K km s$^{-1}$)$^{-1}$.

CO line ratios are typically used to constrain the excitation conditions (temperature, density, and column density) of the 
line-emitting gas.
Some studies have shown that these line ratios (mainly $R_{\rm 21}$) tend to radially vary, 
with higher values toward galaxy nuclei \citep[e.g.,][]{lindt-krieg08,santi09}, 
and this effect presumably reflects similar changing of the physical conditions that drive the 
lower $X_{\rm CO}$ factors found by \citet{sandstrom13}. When presenting a review of the theoretical underpinning, techniques, and results of efforts to estimate 
$X_{\rm CO}$ conversion factor in different environments,
\citet{bolatto13} also stress that $X_{\rm CO}$ appears 
to drop in the central, bright regions of some but not all galaxies.

The AGN may also play a role since a higher $R_{\rm 21}$ toward the center of active galaxies 
could be interpreted as evidence of external heating of molecular clouds by X-rays in the vicinity 
of the AGN \citep[e.g.,][]{baker03}.
However, this radial trend of the gas physical conditions does not unambiguously characterize  
all the LLAGN in our sample. 
Although $R_{21}$ ratios show a radial trend, or hints of it, in NGC~4569, NGC~4579, and NGC~4826, 
$R_{21}$ in NGC~3627 and $R_{32}$ in NGC~4569 and NGC~4579 do not exhibit the same behavior when
assuming a constant or stochastic variation as a function of radius.
For the lack of a common trend in our sample, we adopted constant values of $R_{21}$ and $R_{32}$.
This assumption may introduce an uncertainty into the \sigmahtwo derivation, 
although NUGA galaxies benefit from individual measures of line ratios for each galaxy nuclear region.

As for metallicity, \citet{boissier03} and \citet{narayanan12} find that by using metallicity-dependent $X_{\rm CO}$, 
$\Sigma\rm_{H_{2}}$ should typically change by a factor of 2 to 3 in the central regions compared to disk 
regions of nearby spirals, in the sense of decreasing the gas mass.
Our sample galaxies have all been observed by \citet{moustakas10}, who find
roughly solar metallicities for all but were unable to evaluate metallicity
gradients for lack of data.
Nevertheless, because of the relatively small FoV afforded by our CO observations,
any metallicity gradient is expected to be very slight over the radial range sampled by our data.
We therefore adopted a constant $X_{\rm CO}$ for the sake of uniformity and simplicity, as 
suggested in the review of \citet{bolatto13} in the absence of a detailed characterization 
of $X_{\rm CO}$ in a given galaxy and as typically done in the major portion of studies aimed at determining molecular 
gas mass from $^{12}$CO line intensities.

As already discussed in Sect.~\ref{sec:intro}, we only used molecular (hydrogen) gas in our study based on the assumption 
that SF formation occurs in molecular cloud cores where H$_{2}$ dominates. 
This implies a two-step process of SF where H$_{2}$ clouds are first formed from \hi, 
and stars are then formed from H$_{2}$.
This leads to the hypothesis that H$_{2}$ should be more directly connected with SFR. 
Moreover, we are interested in galactic centers that are high-density regions of galaxies 
and that can be \hi\ deficient \citep[e.g.,][]{bigiel08}.
\citet{haan08} observed the atomic gas for the galaxies in our sample, 
but the \hi\ beam approximately subtends our entire CO FoV, ranging
from $\sim$20\arcsec\ to $\sim$40\arcsec.
With this beam, NGC~4579 has an \hi\ central hole with a surrounding ring, and
this galaxy also has the lowest $\langle$\sigmahtwo$\rangle$ in our sample
(see Table \ref{tab:fits}).
For the other galaxies,
\hi\ column densities \sigmahi$\approx~$10-13\,\msunpc\ for NGC~4569 and NGC~4826,
and $\sim$100\,\msunpc\ for NGC~3627.
In the first two, mean \sigmahtwo\ are 20$-$40 times higher, so neglecting \hi\ will introduce a few percent systematic underestimate of gas surface density;
nevertheless, the \hi\ beam is large so this could be an underestimate.
NGC~3627 has the highest \sigmahtwo\ of our sample (600--700~\msunpc), but the dense
\hi\ gas is still a relatively small fraction.
In any case, as shown by the model predictions of \citet{krumholz09}
in Fig. \ref{fig:nuga-bigiel}, \sigmahtwo\ in the NUGA galaxies
even at 200~pc resolution are in the surface-density regime where H$_2$ is expected
to dominate.

For most of CO data of our sample, it has been possible to combine interferometric data with single-dish ones, allowing the ``missing flux'' problem to be resolved
(see Table \ref{tab:info-co}).
Because in a interferometer two antennas cannot be placed closer than some minimum 
distance ($D_{\rm min}$), signals that are stronger than some value ($\propto$$\lambda/D_{\rm min}$) will be attenuated.
For only interferometric $^{12}$CO maps can the spatially extended sources therefore not be detected, which would
have the effect of underestimating the gas surface density.

\subsection{Derivation of $\Sigma\rm_{SFR}$: diffuse H$\alpha$, [N~{\sc ii}] contamination, and H$\alpha$ extinction}
\label{sec:derivation-ssfr}
Our values of \sigmasfr\ inferred from H$\alpha$ could be overestimated if some fraction of the emission not directly associated with ionization
by massive stars were
diffuse.
\citet{rahman11} find that K-S slopes are steeper, the stronger the fraction of
diffuse emission in the SFR tracer 
\citep[see also][]{liu11}.
However, they also conclude that diffuse emission either in the SFR tracer or 
in CO do not compromise results in the high column-density regions of the
galaxy they studied, NGC~4254.
Since NGC~4254 has peak \sigmahtwo\ $\la$200\,\msunpc, comparable to the mean \sigmahtwo\ 
of our sample galaxies, we do not expect the 
potential overestimate of SFR from diffuse H$\alpha$ emission to be a 
significant problem.

We removed the [N~{\sc ii}] contamination within the filter bandpass using average
[N~{\sc ii}]/H$\alpha$ values available in the literature, as typically done in narrow-band studies.
However, \citet{blanc09} used integral-field spectroscopy to study the spatially-resolved SF relation in NGC~5194,
and checked the validity of the assumption of a constant [N~{\sc ii}]/H$\alpha$ and the bias introduced by it.
They find an increasing [N~{\sc ii}]/H$\alpha$ ratio as one goes to fainter H$\alpha$ fluxes, which is consistent with the nebular emission
in the faintest parts of the galaxy (mainly in the interarm regions) being dominated by the diffuse ionized gas.
Assuming a constant [N~{\sc ii}]/H$\alpha$ may thus
introduce an underestimation of the H$\alpha$ flux (and therefore of the derived SFR) 
in the galaxy's central regions.

In Sections \ref{sec:intro} and \ref{sec:Ha-data}, we stressed that the H$\alpha$ emission lines 
are subject to extinction within the galaxies, which we took into account for our analysis by adopting 
$A_{V}{\rm (int)}$ values determined for each sample galaxy rather than a mean value.
However, we applied this throughout, as suggested
by \citet{kennicutt83}. This is somewhat of an oversimplification because central regions of galaxies are generally 
subject to greater extinction than disk regions.
This may cause the extinction correction and (therefore) the SFR to be underestimated.

\subsection{Choice of the fitting method}
\label{sec:fitting-choice}
A central component of observational investigations is also the quantification of the correlation between two or more
observed quantities.
Typically, linear regression provides estimates of the zero point and slope of the ``best-fit'' regression line between the
observed data. 
Since each adopted methodology for fitting a power law to the data introduces a bias in the results, 
it is important to understand the limitations of common fitting methods.
Usually in the K-S relation studies a linear regression in logarithmic space is performed, 
but methods differ in the treatment of error bars.
Some works \citep[e.g.,][]{kennicutt07,rahman11,momose13} used the FITEXY algorithm \citep[][]{press89}, which
has the advantage of incorporating errors in both the ordinate and abscissa coordinates, 
thus providing robust regression results, although errors must be assumed to be symmetric in logarithmic space, 
which is not always the case.
Other studies, already quoted in Sect.~\ref{sec:fit} and including the present work, adopted the OLS bisector method, 
which returns a bisector line in x-- and y--axes without the errors being taken into account.
Both methods have the disadvantage of not being able to incorporate upper limits in the minimization.

For this reason, \citet{blanc09}
introduced and used a new method for fitting, based on Monte Carlo approach combined with two-dimensional distribution 
comparison techniques, which is not affected by the above issue.
This method has the strong advantage of including the regions not detected in the CO map, 
including those with negative measured fluxes, in addition to 
incorporating the intrinsic scatter in the SF relation as a free parameter and performing the fitting 
in linear space, thereby avoiding the assumption of log-normal symmetric errors.   
For a detailed discussion of the disadvantages of the linear regressions with respect to 
a Monte Carlo fitting method for a more realistic treatment of systematics and accurate determination of the
parameters and errors, we refer the reader to \citet{blanc09} and \citet{shetty13}.
Such an analysis is beyond the scope of this paper. 
 
Based on the IRAM PdBI CO data reduction performed for all NUGA galaxies, we preferred 
to take only data above 3$\sigma$ level into account, 
for which we are certain of the interferometric significance.
When analyzing the SF relation at sub-kpc scales of ten nearby
galaxies, \citet{momose13} note that 
the choice of CO threshold surface density (3$\sigma$ also in their case) affects the results very little; however . when
repeating the analysis with various thresholds, they obtained the same findings in terms of the K-S relation.
Since we had compared our results with \citet{bigiel08} (see Fig.~\ref{fig:nuga-bigiel}) who used the OLS bisector method,
we also adopted it.

We note that results derived from the robust regression-fitting method can differ very much from OLS bisector ones 
(see Table~\ref{tab:fits}), confirming that the choice of the fitting method affects the findings \citep[][]{blanc09,shetty13}.
This further reinforces the need to compare K-S relation results obtained based on the same fitting method. 


\section{Conclusions}
\label{sec:conclusion}

We have investigated the resolved molecular K-S relation in the central kpc of four NUGA galaxies
that have different types, which interact or do or not, which are barred or not, which have either a Seyfert or a Liner nucleus.
Their SFR ranges from $\sim$0.7 to $\sim$4~M$_{\odot}$~yr$^{-1}$, and sSFR from 0.04 to 0.5~Gyr$^{-1}$.
The spatial scales studied in this work lie between $\sim$20 and 200~pc, and densities sampled
are higher than in the usual galaxy disks, extending toward 100--1000~\msunpc. 

The first result was that each galaxy has its own molecular SF relation on the investigated spatial scales.
The K-S relation is most often underlinear (see NGC~4569), with slopes ranging from $\sim$0.5 to $\sim$1.3.
The derived depletion time scales range between 1 and 2~Gyr, which is very compatible with
what is found on a larger scale in the COLD GASS sample of nearby galaxies.
These results are valid regardless of the $^{12}$CO line observed, from the (1--0) to (3--2) transition.
We also found that the K-S relation changes behavior on spatial scales of $\sim$80--100~pc, with the index $N_{\rm fit}$ 
starting to sharply decrease with finer resolution with respect to larger scales.  
A break in the relation between \sigmahtwo\ and \sigmasfr\ is expected on small scales ($\sim$10--50~pc),
owing to phenomena such as star formation feedback, the life time of clouds, a turbulent cascade, or magnetic fields 
\citep[e.g.,][]{kruijssen14}, and it has been observed in M~51 \citep[][]{liu11} and M~33 \citep[][]{onodera10} in dense nuclei.
However, in dense nuclei available thanks to NUGA sample a net break is still not visible on scales of 20 pc.
This might be due to the higher density of the GMCs present in galaxy centers,
which have to resist higher shear forces.
Finally, one of the most important findings was that the proportionality between \sigmahtwo\ and \sigmasfr\ found
between 10 and 100~\msunpc\  continues at densities as high as $\sim$10$^{3}$~\msunpc.
However, with a common conversion factor, the central regions of the NUGA galaxies have higher
\sigmasfr\ for a given gas column than those expected from the models, with behavior 
between the mergers or high-$z$ starburst systems and the more quiescent SFGs,  
assuming that the former require a lower value of $\alpha_{\rm CO}$.

\begin{acknowledgements}
We thank the anonymous referee for useful comments and suggestions that improved the quality 
of the manuscript. 
We also thank Daniela Calzetti for precious comments and suggestions.
V. Casasola also wishes to thank Laura Magrini and Jacopo Fritz for helpful discussions on star-formation processes.
This research made use of the NASA/IPAC Extragalactic Database (NED), which is operated by the Jet Propulsion Laboratory, 
California Institute of Technology, under contract with the National Aeronautics and Space Administration. 
This research also made use of observations with the NASA/ESA Hubble Space Telescope
and others obtained from the Hubble Legacy Archive, which is a collaboration between the Space Telescope 
Science Institute (STScI/NASA), the Space Telescope European Coordinating Facility (ST-ECF/ESA),
and the Canadian Astronomy Data Centre (CADC/NRC/CSA).
\end{acknowledgements}


\end{document}